\documentstyle[preprint,aps,psfig]{revtex}

\newcommand{\nc}{\newcommand}

\nc{\dbar}{\bar{\partial}}
\nc{\be}{\begin{equation}}
\nc{\ee}{\end{equation}}

\nc{\beq}{\begin{equation}}
\nc{\eeq}{\end{equation}}
\nc{\bea}{\begin{eqnarray}}
\nc{\eea}{\end{eqnarray}}
\nc{\beqa}{\begin{eqnarray}}
\nc{\eeqa}{\end{eqnarray}}
\nc{\nn}{\nonumber}
\newcommand{\al}{\alpha^{\prime}}

\catcode`\@=11

\@addtoreset{equation}{section}
\def\theequation{\arabic{section}{.}\arabic{equation}}
\@twosidetrue
\relax

\catcode`@=12

\begin{document}

\begin{titlepage}

\begin{flushright}

SISSA REF 11/98/EP

\end{flushright}

\vskip 10pt

\begin{center}

{\Large\bf Scale independent spin effects in \\

\vskip 5pt

D-brane dynamics}
 
\vskip 20pt

{\large Jose F. Morales$^{a}$, Claudio A. Scrucca$^{a,b}$ and 
Marco Serone$^{c}$}

\vskip 18pt

$^{a}$International School for Advanced Studies, ISAS-SISSA, \\
Via Beirut 2-4, 34013 Trieste, Italy

\vskip 2pt

$^{b}$Istituto Nazionale di Fisica Nucleare, Sez. di Trieste, Italy

\vskip 2pt

$^{c}$ Dept. of Mathematics, University of Amsterdam, 
Plantage Muidergracht 24, \\ 
1018 TV Amsterdam, The Netherlands

\vskip 2pt

e-mail: morales,scrucca@sissa.it; serone@wins.uva.nl 

\vskip 5pt
 
\end{center} 

\begin{abstract}
We study spin interactions between two moving D-branes using the 
Green-Schwarz formalism of boundary states. 
We focus our attention on the leading terms for
small velocities $v$, of the form $v^{4-n}/r^{7-p+n}$ ($v^{2-n}/r^{3-p+n}$) 
for p-p (p-p+4) systems, with 16 (8) supercharges. 
In analogy with standard G-S computations of massless four-point 
one-loop amplitudes in Type I theory, the above terms are 
governed purely by zero modes, massive states contributions 
cancelling as expected by the residual supersymmetry.
This implies the scale invariance of these leading spin-effects,
supporting the relevant matrix model descriptions of supergravity
interactions; in this context, we also discuss similar results for 
more general brane configurations.
We then give a field theory interpretation of our results, that
allows in particular to deduce the gyromagnetic ratio $g=1$ and the presence
of a quadrupole moment for D0-branes.
\end{abstract}
 
\vskip 10pt
\begin{flushleft}
\rule{16.1cm}{0.2mm}\\[-3mm]
\end{flushleft}
\begin{flushleft}
\end{flushleft}

\end{titlepage}

\section{Introduction}

It is undoubt that D-branes \cite{pol} play at present a key role in the 
description of all non-perturbative phenomena in string theory. This has 
motivated the study of their dynamics \cite{bach,dan,kp,lif,dkps} in different 
configurations and space-time dimensions; in particular, the so-called boundary
state formalism \cite{pocai,clny} has proven to be a powerful tool for 
studing these problems, and has been succesfully applied to a number of 
problems, both in the covariant 
\cite{li,calkl,billo,lerda1,lerda2,hin,hins,bis}
and light-cone \cite{gre,gregut1,gregut2,barb} formulations.
However, since D-branes are usually analyzed in a semi-classical 
context, being heavy massive solitons at weak string coupling constant, most 
of these works considered their interactions in the approximation in which 
D-branes do not carry any spin.

The conjecture of \cite{bfss} that the dynamics of M-theory in the infinite 
momentum frame is governed by the degrees of freedom of a large number of 
D0-branes further motivated the study of D0-brane dynamics. In particular, 
their spin-dependent long-range interactions have been computed through
duality arguments, mapping the scattering of D0-branes to that of 
fundametal KK states \cite{harv}. 
In ref.\cite{noi}, spin effects for generic p-branes have been 
analyzed in the Green-Schwarz formalism, by inserting broken supercharges in the 
partition function of two branes moving past each other. The structure of these 
interactions was then considered in a long range regime through the technique of 
boundary states and at short distances through a one-loop annulus analysis. 
This last open string point of view showed, in particular, how for a D p-brane 
all the terms of the form $v^{4-n}/r^{7-p+n}$ with $n=0,...,4$, i.e. all the 
spin-effects associated to the universal $v^4/r^{7-p}$ interaction, are determined 
purely by the {\it open} massless string states, meaning then that a one-loop
M(atrix) theory computation should be able to reproduce long range spin-dependent
supergravity interactions. This has been indeed explicitly shown in \cite{kra} for 
the spin-orbit $v^3/r^8$ coupling.

We want then to study interactions between two branes moving with small relative 
velocity $\vec{v}$ and at an impact parameter $\vec{b}$. 
We will work in the Green-Schwarz formalism of the boundary state. The 
choice of this formalism is justified by the huge simplification that takes over 
and that allows to compute spin effects basically by a simple analysis of zero 
modes, as we will see. We consider in the 
following supersymmetric configurations associated to the case of parallel p-p and  
p-p+4 systems of D-branes. 
Being interested only in leading effects for small velocity, we will not compute the 
full partition function associated to the configuration of two moving branes, but 
simply correlation functions of vertex operators, corresponding to the velocity, 
on the boundary of the world-sheet, together with insertions of broken 
supercharges encoding spin dependences. The approach we follow to 
compute $n$-point functions in the boundary state formalism is that derived by 
\cite{pocai,clny}. It is crucial to point out, however, that the results we 
obtain are {\it exact} for the particular terms considered, being given by a full 
string computation, in analogy with one-loop 4-point functions between massless 
states of Type I theory in the G-S formalism \cite{gsw}. 
In this context it is trivial to see some important 
issues.  A p-p (p-p+4) system leaves unbroken sixteen (eight) supercharges meaning 
that our action will admit eight (four) fermionic zero modes $S_0^a$, in the 
notation of \cite{gsw}. Since, as we will see, each insertion of a vertex operator 
associated to the velocity provides at most two of them, it follows that the potential 
between branes 
starts respectively with $v^4$ ($v^2$). On the other hand, the insertion of broken
supercharges can allow non-vanishing results for terms with less powers of $v$, 
providing the lacking fermionic zero modes. Moreover, as has been shown in 
\cite{gregut2,noi} and will be seen again in some detail, the insertion of 
supercharges induces polarization-dependent non-minimal couplings between
D-branes, i.e. spin-effects. Alternatively, being all these terms related by 
supersymmetry \cite{harv}, it is natural that they are produced by insertion of 
supercharges.

The main point we want to stress here is that for the amplitudes we consider, 
associated to the $v^{4-n}/r^{7-p+n}$ terms discussed before
\footnote{For the p-p+4 system the corresponding terms are of the form 
$v^{2-n}/r^{3-p+n}$, with $n=0,1,2$.}, our results will be governed by zero modes 
only, since massive non-BPS bosonic and fermionic string states contribution will always 
precisely cancel. This implies that the present results, that for large brane 
separations have a clear interpretation as spin-dependent interactions, due to 
supersymmetry, are valid at {\it all scales} and can be extrapolated to the 
substringy regime where, according to \cite{dkps}, the dominant degrees of freedom
are given by the massless open string states living on the branes. 
This assertion is valid for a large class of interesting systems
discussed in the final section, besides the more detailed studied
p-p and p-p+4 system, and shows that several tests (if not all)
performed until now to the proposal of \cite{bfss}, involving 
an agreement between one-loop Super Yang-Mills computation and classical 
supergravity, are basically determined by supersymmetry.

We then present a systematic way to construct the generic asymptotic form of
spinning p-brane supergravity solutions, working out in detail the case of the 
D0-brane, where we perform a multi-pole expansion up to the quadrupole interaction.
This analysis allows easily to show that the electric and gravitational dipole
moments for D0-branes vanish, whereas there is a gravitational quadrupole 
term different from zero.
The supersymmetric cancellation of some terms of brane-interactions, as discussed
before, allows finally to fix uniquely the value of D0-brane's gyromagnetic ratio 
$g$ to one and similarly the relative strength of the electric and gravitational
quadrupole moments. This is consistent
with their conjectured \cite{witten} 11-dimensional KK origin and interpretation 
as solitonic solution of 10-dimensional IIA supergravity, as recently discussed in 
ref.\cite{dlr}. As a further check of this correspondence, it would be 
interesting to work out, along the lines of \cite{dlr}, the quadrupole moments of 
0-brane solitons in supergravity and compare them to the results found here for D0 branes.
We then briefly mention how the knowledge of the asymptotic fields of the 0-brane
solution allows to derive the 0-brane world-line effective action in an arbitrary
Type IIA background, valid in the weak-field approximation.

The structure of this paper is as follows. In section 2 and 3 we review the main
properties of G-S boundary states \cite{gregut1}, introduce the set-up of the
computation, and report one-point functions of higher spin boundary states with
massless closed string states. In section 4 and 5, we compute spin-dependent interactions
associated to the p-p and p-p+4 parallel brane configurations, and in section 6 the field
theory interpretation of our results is given.
In last section we discuss possible extensions of our results to more general
brane configurations, relevant for one-loop tests of the matrix model conjecture,
and give some conclusions. 
We include in an appendix 
the light-cone computations of the standard phase-shift for two moving D-branes. 

\section{Boundary state formalism}

In this section we shall briefly review the boundary state description 
of spinning D-branes \cite{noi} in the G-S formalism \cite{gregut1}. 
Consider type II theory in the light-cone gauge $X^+ = x^+ + p^+ \tau$. 
$X^-$ is completely determined in terms of the transverse $X^i$'s, in
${\bf 8_v}$, and the left and right spinors $S^a$ and $\tilde S^a$, 
in the ${\bf 8_s}$ of the $SO(8)$ transverse rotation group 
\footnote{In the following we shall concentrate on Type IIB theory, for which
the notation is somewhat frendlier; the Type IIA case can be easily obtained 
by switching dotted and undotted indices in the right-moving fermions.}.
In this gauge, the supercharges are 
$$
Q^a = \sqrt{2 p^+} \oint d\sigma S^a \;,\;
Q^{\dot a} = \frac 1{\sqrt{p^+}} \gamma^i_{\dot a a}
\oint d\sigma \partial X^i S^a 
$$
with similar expressions for the right moving ones.

Dp-branes as defects can be described by suitable boundary states,
implementing the usual Neumann-Dirichlet boundary conditions.
In this frame, the two light-cone directions $\pm = 0\pm 9$
satisfy automatically Dirichlet boundary conditions while the transverse
directions $i=1,...,8$ can have either Neumann or Dirichlet boundary conditions.
Since the time satisfies Dirichlet boundary conditions, 
we are actually dealing with Euclidean branes;
however following \cite{noi}, we can identify the ``time'' with one of the transverse
directions, say $X^1$. The usual metric is then recovered
through a double analytic continuation $0 \rightarrow i1$, 
$1 \rightarrow i 0$ in the final result.
We can therefore include in our analysis only branes with 
$p=-1,..., 7$. The BPS boundary state is defined to preserve a combination of left 
and right supersymmetries 
\beqa
Q^a_+|B\rangle = 0  &\;,\;& Q^{\dot a}_+ |B\rangle = 0 \;\Rightarrow\; 
Q^a_+,Q^{\dot a}_+ \;\mbox{unbroken} \nn \\
Q^a_-|B\rangle \neq 0  &\;,\;& Q^{\dot a}_- |B\rangle \neq 0 \;\Rightarrow\; 
Q^a_-,Q^{\dot a}_- \;\mbox{broken}\nn
\eeqa  
where 
$$
Q^a_\pm = \frac 1{\sqrt{2}} \left(Q^a \pm i M_{ab}\tilde Q^b \right) \;,\;\;
Q^{\dot a}_\pm = \frac 1{\sqrt{2}} \left(Q^{\dot a} 
\pm i M_{\dot a \dot b} \tilde Q^{\dot b} \right)
$$
The boundary conditions are \cite{gregut1}
\be
\label{bccc}
(\alpha^i_n+M_{ij}\tilde{\alpha}^j_{-n})|B\rangle=0 \;,\;\;
(S^a_n + i M_{ab} \tilde S^b_{-n} )|B\rangle=0\;,\;\;
(S^{\dot a}_n + i M_{\dot a \dot b} \tilde S^{\dot b}_{-n} )|B\rangle=0
\ee
with
\be
M_{ij} = \pmatrix {- I_{p+1}& 0 \cr 0 & I_{7-p} \cr}
\label{Mv} 
\ee
Consistency with the BPS condition implies the $M$'s to be orthogonal 
matrices related by triality
\beqa
&&(MM^T)_{ab} = \delta_{ab} \nn \\
&&(M \gamma^i M^T)_{a \dot a} = M_{ij} \gamma^j_{a \dot a} \nn
\eeqa
which yields
\be
M_{ab} = (\gamma^1 \gamma^2 ... \gamma^{p+1})_{ab} \;,\;
M_{\dot a \dot b} = (\gamma^1 \gamma^2 ... \gamma^{p+1})_{\dot a \dot b}
\label{Ms}
\ee

The solution for the boundary state can then be found in a standard way as the 
eigenstate of the boundary conditions eqs.(\ref{bccc}), and is written as
\be
\label{mom}
|B\rangle =\exp\sum_{n>0}\left({1\over n}  M_{ij}
\alpha^i_{-n}\tilde{\alpha}^j_{-n} -
i M_{ab}S^a_{-n}\tilde{S}^{b}_{-n}\right)|B_0\rangle
\ee \noindent
the zero mode part being
\be
\label{zm}
|B_0\rangle= M_{ij} |i\rangle|\tilde j\rangle 
- i M_{\dot a \dot b}|\dot a\rangle|\tilde{\dot b}\rangle \label{zero}
\ee \noindent
The complete configuration space boundary state is
\be
\label{conf} 
|B,\vec{x}\rangle=(2\pi\sqrt{\al})^{4-p}\int\frac{d^{9-p}q}{(2\pi)^{9-p}}
\,e^{i \vec{q}\cdot\vec{x}}\,|B\rangle\otimes|\vec{q}\rangle
\ee \noindent
where
$$
\langle q|q^{\prime}\rangle =V_{p+1}\,(2\pi)^{9-p}
\delta^{(9-p)}(q-q^{\prime}) 
$$
and $V_{p+1}$ is the space-time volume spanned by the p-brane. 
With these normalizations, the static force between two parallel branes is given by 
\begin{equation}
{\cal A}=\frac{1}{16}\int_0^\infty \!\!dt \,
\langle B,\vec{x}|e^{-2\pi t\al p^+(P^--i\partial/\partial x^+)}| B,\vec{y} \rangle
\label{cyli}
\end{equation}
where
$$
P^-=\frac{1}{2p^+}\left[(p^i)^2+\frac{1}{\al}\sum_{n=1}^\infty (
\alpha^i_{-n} \alpha^i_{n}+\tilde{\alpha}^i_{-n} \tilde{\alpha}^i_{n}  
+ n \,S^a_{-n} S^a_{n}+ n \,\tilde{S}^{a}_{-n}\tilde{S}^{a}_{-n}
)\right] 
$$
is the Hamiltonian in the light-cone gauge.
The term $i\partial_+$ represents the substraction of $p^-$ (remember that in
this gauge the effective Hamiltonian is $H-p^-$) and when applied to the 
boundary state, it reproduces simply the covariant $p^2$. The factor 1/16 
is needed to normalize correctly the D-brane charge; indeed, from eq.(\ref{cyli}) 
we obtain
\be 
{\cal A}=\frac{1}{16}V_{p+1}\,(4\pi^2\al)^{4-p}\int_0^\infty \! dt 
\int\frac{d^{9-p}q}{(2\pi)^{9-p}}\,e^{i \vec{q}\cdot (\vec{x}-\vec{y})}
\,e^{-\pi t\al q^2} (8-8)\prod_{n=1}^{\infty}\frac{(1-e^{-2\pi tn})^8}
{(1-e^{-2\pi tn})^8} \label{static}
\ee
where the factor $(8-8)$ is due to the trace performed on the zero mode part of the
boundary state, eq.(\ref{zero}). Performing the momenta and modulus integrations, 
one finds
\be
{\cal A}=2\,V_{p+1}\,T_p^2\,G_{9-p}(\vec{x}-\vec{y})\,(1-1) \label{1-1}
\ee\noindent
where $T_p=\sqrt{\pi}(4\pi^2\al)^{(3-p)/2}$ is the tension of a p-brane in units of 
the ten-dimensional Planck constant $k^2$ of Type II supergravity \cite{pol} and
$G_d(\vec{x})$ is the massless propagator of a scalar particle in $d$-dimensions
$$ 
G_d(\vec{x})=\frac{1}{4\pi^{d/2}}\frac{\Gamma(\frac{d-2}{2})}{|x|^{d-2}} 
$$
The generalization of the cylinder amplitude to the case of finite constant
relative velocity $v$ between two branes is reported for completeness in the appendix.

\section{Supersymmetry and higher spin BPS states}

As we have seen, Dp-branes correspond to solitonic BPS saturated solutions of 
Type IIA(B) supergravity, which preserve one half of the supersymmetries.
The remaining half is realized on a short-multiplet containing 256 p-brane 
configurations lying in the {\bf 44}+{\bf 84}+{\bf 128} representations of the little 
group $SO(9)$ for massive states. 
The various components of the short-multiplet are related by 
supersymmetry transformations generated by the 16 broken supercharges.

In the formalism of previous section, the boundary state represents
the semiclassical source formed by the ``in'' and ``out'' branes; its 
overlap $\langle B|\Psi\rangle$ with a string state $|\Psi \rangle$ represents
semiclassical 3-point functions as shown in figure 1.

\vskip 30pt

\input epsf
\epsfsize=20pt
\centerline{\qquad \qquad \quad \epsffile{0q.eps}}
\vskip -56pt
$\qquad \qquad \qquad \qquad \qquad \quad \langle \Psi_B|B \rangle = 
\qquad \qquad \qquad \qquad \quad \Psi_B$
\vskip 56pt
\vskip -115pt
$\qquad \qquad \qquad \qquad \qquad \qquad \qquad \qquad B_B$
\vskip 115pt
\vskip -59pt
$\qquad \qquad \qquad \qquad \qquad \qquad \qquad \qquad B_B$
\vskip 59pt
\vskip -30pt

\input epsf
\epsfsize=20pt
\centerline{\qquad \qquad \quad \epsffile{1q.eps}}
\vskip -56pt
$\qquad \qquad \qquad \qquad \quad \; \langle \Psi_F|Q^-|B \rangle =
\qquad \qquad \qquad \qquad \quad \Psi_F$
\vskip 56pt
\vskip -115pt
$\qquad \qquad \qquad \qquad \qquad \qquad \qquad \qquad B_B$
\vskip 115pt
\vskip -59pt
$\qquad \qquad \qquad \qquad \qquad \qquad \qquad \qquad B_F$
\vskip 59pt
\vskip -30pt

\input epsf
\epsfsize=20pt
\centerline{\qquad \qquad \quad \epsffile{2q.eps}}
\vskip -56pt
$\qquad \qquad \qquad \quad \;\;\, \langle \Psi_B|Q^-Q^-|B \rangle =
\qquad \qquad \qquad \qquad \quad \Psi_B$
\vskip 56 pt
\vskip -115pt
$\qquad \qquad \qquad \qquad \qquad \qquad \qquad \qquad B_F$
\vskip 115pt
\vskip -59pt
$\qquad \qquad \qquad \qquad \qquad \qquad \qquad \qquad B_F$
\vskip 59pt

\vskip -40pt

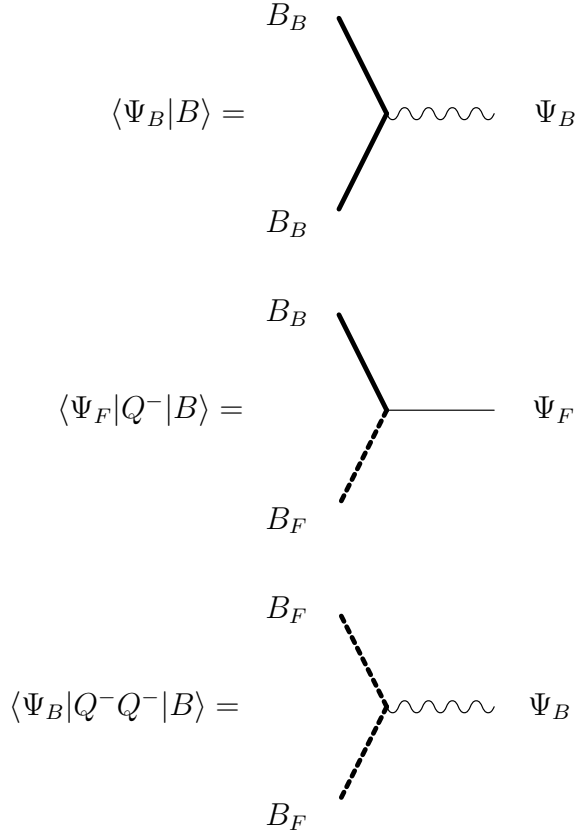
\centerline{\bf Fig. 1}

\noindent
Different components of the supermultiplet spanned by these sources, are obtained by 
applying supersymmetry transformations to the scalar boundary state $|B\rangle$
through the operator
\be
\label{u}
U = e^{\epsilon Q^-} = \sum_{m=0}^{16} \frac 1{m!} (\epsilon Q^-)^m 
\ee
We have used the $SO(9)$ notation $\epsilon = (\eta^a, \tilde \eta^{\dot a})$ 
and $Q^- = (Q_-^a,Q_-^{\dot a})$. 
Different components of the supermultiplet, corresponding to the possible independent 
$\epsilon$'s, can be thought as the semiclassical multipole expansion of the source.
In particular, terms in $U|B\rangle$ with an even (odd) number of $Q^-$ describe
couplings to bosonic (fermionic) closed string states $\Psi_B$ ($\Psi_F$).   
We shall restrict ourselves to terms with an even number of supercharges, the 
relevant for the study of semiclassical D-brane dynamics in the eikonal approximation.
For instance, the usual boundary state corresponds to the universal
part of the source, whereas applying two charges one obtains the part of the source due
to angular momentum, and so on. As we are going to see in the following, the field theory
counterpart of this source expansion is a power series in the transfered momentum, each 
momentum corresponding to the insertion of a pair of supercharges.  

\noindent
Among the different terms in expansion (\ref{u}) we will always work out 
the ones with an equal number of dotted-undotted $SO(8)$ components
$(\eta_a Q_-^a\tilde \eta_{\dot a} Q_-^{\dot a})^n$. All the other terms
simply combine to reconstruct the covariant answer. We consider then 
boundary states of the form:
\be
|B\rangle_{(n)}\equiv V_{\eta}^n |B\rangle
\label{bn}
\ee
with
\be
V_{\eta} = \eta_a Q^{-a} \tilde \eta_{\dot a} Q^{-\dot a}
\label{veta} 
\ee
The first interesting information we can extract from these higher spin 
boundary states is about their couplings to the massless bulk fields. 
This analysis for the D-instanton case was performed in the covariant NS-R formalism 
in ref.\cite{gregut2}. The formulae displayed in this section are ``T-dual''
versions of the ones reported in that reference.

\vskip 20pt

\input epsf
\epsfsize=50pt
\centerline{\epsffile{1p.eps}}
\vskip 10pt
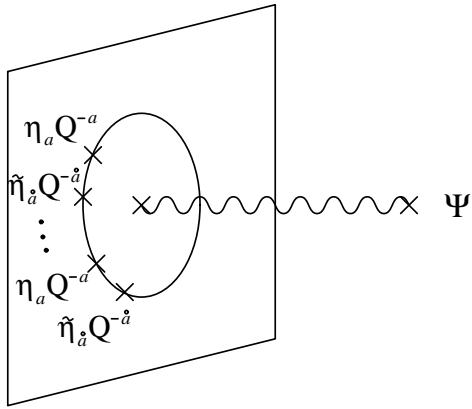
\centerline{\bf Fig. 2}

\vskip 10pt

In the following, we consider only terms with up to eight supercharges, $n=0,...,4$, 
in eq.(\ref{bn}). This covers all the physical information relevant to our considerations.
The one-point functions of the massless bosonic states of NSNS and RR sectors 
(in R-NS terminology) are obtained simply by computing the boundary state overlap
\beq
\Psi_{(n)} \equiv \langle \Psi|B_0\rangle_{(n)}
\label{onepoint}
\eeq
with the massless NSNS and RR closed string states
\beqa
|\Psi_{NSNS}\rangle &=& \xi_{mn}|m\rangle \tilde{|n\rangle} \;\Rightarrow\;
\xi_{mn}\sim\phi \,\delta_{mn} + g_{mn} + b_{mn} \nn \\
|\Psi_{RR}\rangle &=& C_{\dot a \dot b}|\dot a\rangle \tilde{|\dot b\rangle}
\;\Rightarrow\; C_{\dot a \dot b} \sim \sum_{k~even} \frac 1{k!} 
C_{(k)}^{i_1 ... i_k} \gamma^{i_1 ... i_k}_{\dot a \dot b} \nn
\eeqa
$|B_0\rangle_{(n)}$ indicates the massless content of (\ref{bn})
\beq
|B_0\rangle_{(n)} \equiv
V^n_{\eta}|B_0\rangle = q_{i_1} ... q_{i_n}
\left[\eta_{[a_1} (\tilde \eta \gamma^{i_1})_{a_2} 
... \eta_{a_{2n-1}} (\tilde \eta \gamma^{i_n})_{a_{2n}]}\right]
S_0^{-a_1}...S_0^{-a_{2n}}|B_0\rangle
\label{Vn0}
\eeq
Using the boundary conditions 
eqs.(\ref{bccc}), we can express everything in terms of the left-moving modes 
only. After applying the Fiertz identity
$$
S_0^{a}S_0^{b}= \frac 12 \delta^{ab} + \frac{1}{4} \gamma^{ij}_{ab} R_0^{ij}
$$ 
we can further rewrite eq.(\ref{Vn0}) in terms of the $SO(8)$ generators
$R_0^{ij}=1/4S_0^a\gamma^{ij}_{ab}S_0^b$. 
We are then left with the effective operator
\be
V_{\eta}^n = q_{i_1} ... q_{i_n} \,
\omega^{i_1...i_{n}}_{j_1...j_{2n}}(\eta)
R_0^{j_1 j_2} ... R_0^{j_{2n-1} j_{2n}}\;,
\ee
where
\be
\omega_{j_1...j_{2n}}^{i_1...i_{n}}(\eta) =
\frac 1{2^n} \left[\eta_{[a_1} (\tilde \eta \gamma^{i_1})_{a_2} 
... \eta_{a_{2n-1}} (\tilde \eta \gamma^{i_n})_{a_{2n}]}\right]
\gamma^{j_1 j_2}_{a_1a_2} ... \gamma^{j_{2n-1} j_{2n}}_{a_{2n-1} a_{2n}}
\ee
encodes the spin dependence.
In this way, we can use standard results for Type I theory. The $R_0^{ij}$ 
generators are represented in the ${\bf 8_v}$ and ${\bf 8_c}$ representations by
\bea
&&R_0^{mn}|i\rangle = (\delta^{ni} \delta^{mj}-\delta^{mi} \delta^{nj})|j\rangle \nn \\
&&R_0^{mn}|\dot a\rangle = \frac 12 \gamma^{mn}_{\dot c \dot a} |\dot c\rangle 
\label{R0} 
\eea
and some simple algebra leads to 
$$
|B_0\rangle_{(n)} \equiv M_{(n)}^{ij} |i\rangle \tilde{|j\rangle}
- i M^{(n)}_{\dot a \dot b} |\dot{a}\rangle\tilde{|\dot b\rangle}
$$
with
\bea
M_{ij}^{(n)} &=& 2^n \,q_{i_1} ... q_{i_n} \,
\omega^{i_1 ... i_n}_{ik_1k_1...k_{n-1}k_{n-1}k_n}(\eta)M_{k_nj} \\
M^{(n)}_{\dot a \dot b} &=& \frac 1{2^{n}} \, 
q_{i_1} ... q_{i_n} \, \omega^{i_1 ... i_n}_{j_1...j_{2n}}(\eta)
(\gamma^{j_1 j_2} ... \gamma^{j_{2n-1} j_{2n}}M)_{\dot{a}\dot{b}} 
\eea
The 1-point functions can then be written as (up to numerical and ${\al}$ factors) 
\bea
\Psi^{NSNS}_{(n)} &=& q_{i_1} ... q_{i_n} \,
\xi^{ij} \omega^{i_1 ... i_n}_{i k_1 k_1 ... k_{n-1}k_{n-1}k_n}(\eta) 
M_{k_n j} \label{NS} \\
\Psi^{RR}_{(n)} &=& q_{i_1} ... q_{i_n} \,
\sum_{k} \frac 1{k!} C_{(k)}^{m_1 ... m_k} 
\omega^{i_1 ... i_n}_{j_1 ... j_{2n}}(\eta)
\mbox{Tr}_S [\gamma^{m_1 ... m_k}\gamma^{j_1 j_2} ... \gamma^{j_{2n-1}j_{2n}}M]
\label{R}
\eea
Eqs.(\ref{NS}) and (\ref{R}) contain all the non-minimal couplings of D-branes 
with the massless bosonic states of the corresponding supergravity theory.
In particular, for even $n$ the boundary state couples potentially to the NSNS components
$\phi,g_{\mu\nu}$, $g_{IJ}$ and $b_{\mu I}$ ($\mu,\nu$ and $I,J$ denoting
Neumann and Dirichlet directions respectively), and to the remainig NSNS fields for 
odd $n$, as can be seen using the symmetry properties of $\omega^{i_1 ...  i_{2n}}$. 
As a source of RR fields, we can see that non-minimal couplings arise from the 
non-vanishing gamma-traces in eq.(\ref{R}), corresponding to forms with 
$k=p+1-2n, ... ,p+1+2n$. The specific form of these couplings depends 
on the polarization details and will be explicited for the first terms 
in the following.    

The first universal NSNS coupling is just
\be 
\Psi^{NSNS}_{(0)}=\xi_{ij}M_{ji} \label{NS0} 
\ee 
We see that any p-brane couples to 
a specific linear combination of the dilaton $\phi$ and the diagonal graviton polarizations 
$g_{11} ... g_{p+1,p+1}$, as it must be for an object with definite mass density
\footnote{The only exception is the D-instanton that has at this order only the 
coupling to the dilaton ($M_{ij}=\delta_{ij}$).} 
(remember that $g_{11}\rightarrow -g_{00}$ after analytic continuation). 
The RR coupling is
\be
\Psi^{RR}_{(0)} =\sum_{k~even} \frac 1{k!} C_{(k)}^{i_1 ... i_k} 
\mbox{Tr}_S [\gamma^{i_1 ... i_k}M] \label{RR0}
\ee
The gamma-trace vanishes unless $k=p+1$, giving the usual Dp-brane charge.
The covariant expressions of eqs.(\ref{NS0}) and (\ref{RR0}) are
\be
\Psi^{NSNS}_{(0)}=\xi_{\mu\nu}M^{\nu\mu} \;,\;\;
\Psi^{RR}_{(0)} =\sum_{k~even} \frac 1{k!} C^{(k)}_{\mu_1 ... \mu_k} 
\mbox{Tr}_S [\Gamma^{\mu_1 ... \mu_k}M] \label{cov}
\ee
where $\Gamma$ are $SO(1,9)$ gamma-matrices, $M^{\nu\mu}$ is the covariant 
extension of eq.(\ref{Mv}), with diagonal entries only, -1 and +1 in Neumann and Dirichlet 
directions respectively, and ${\cal M}=\Gamma^0...\Gamma^{p}$.

The first non-minimal NSNS coupling is given by
\be 
\label{NS2} 
\Psi^{NSNS}_{(1)} = \xi_{ij} M_{jk} q_l \omega_{ki}^l
=\xi_{ij} M_{jk}\,q_l \,\eta\gamma^{kil}\tilde{\eta}  
\ee
where we have used the fact that $q_j\xi_{ij}=q_i\xi_{ij}=0$ and $q_kM_{kj}=q_j$ 
(there is a non-vanishing momentum transfer only along the Dirichlet directions). 
As anticipated, eq.(\ref{NS2}) represents a non-minimal coupling of the brane with the 
antisymmetric tensor and graviton polarizations $b_{\mu\nu}$, $b_{IJ}$ and $g_{\mu I}$.
The covariant expression of eq.(\ref{NS2}) is simply
\be 
\Psi^{NSNS}_{(1)} = \xi_{\mu\sigma}M^{\sigma}_{\nu}q_{\rho}\bar{\psi}\Gamma^{\mu\nu\rho}\psi
\label{cNS1} 
\ee
where $\psi$ is the Majorana-Weyl fermionic parameter associated to the broken 
supersymmetry. In a chiral representation, it is simply 
$\psi=\left(^{\textstyle \epsilon}_0\right)$, where 
$\epsilon = (\eta^a, \tilde \eta^{\dot a})$. 
The corresponding RR coupling is
\be 
\Psi^{RR}_{(1)} =\sum_{k~even} \frac 1{k!} C_{(k)}^{i_1 ... i_k} 
\mbox{Tr}_S (\gamma^{i_1 ... i_k}\gamma^{ij}M) 
\,q_l \,\eta\gamma^{ijl}\tilde{\eta} 
\label{R2} 
\ee
where still the completely antisymmetric part in the fermion bilinear is the
only non-vanishing contribution, since $q^{i_1} C_{(k)}^{i_1 ... i_k} =0$.
The covariant form of eq.(\ref{R2}) is
\be 
\Psi^{RR}_{(1)} =\sum_{k~even} \frac 1{k!} C^{(k)}_{\mu_1...\mu_k}
\mbox{Tr}_S (\Gamma^{\mu_1...\mu_k}\Gamma_{\nu\rho}{\cal M})q_{\sigma}
\bar{\psi}\Gamma^{\nu\rho\sigma}\psi \label{cR1} 
\ee

The next coupling is $\Psi^{NSNS}_{(2)}$, that is
\be 
\Psi^{NSNS}_{(2)} = \xi_{ij}\,q_{j_1} q_{j_2}\omega_{ii_1i_1i_2}^{j_1j_2}
M^{i_2j}
\label{NS4} 
\ee
After some algebra eq.(\ref{NS4}) can be rewritten, neglecting $q^2$ contact terms 
which are irrelevant for our semiclassical analysis, as
\be 
\Psi^{NSNS}_{(2)} =\tilde{\xi}_{ij}\,q_mq_n
\,(\eta\gamma^{ink}\tilde{\eta}\,\eta\gamma^{jmk}\tilde{\eta}-
\eta\gamma^{im}\eta\,\tilde\eta\gamma^{jn}\tilde\eta) \label{NSS4} 
\ee
where $\tilde{\xi}_{ij}\equiv\xi_{ik}M_{kj}+\xi_{jk}M_{ki}$. Notice that the 
combination of spinors in eq.(\ref{NSS4}) is the right one reproducing the 
covariant expression \cite{gregut2}
\be 
\Psi^{NSNS}_{(2)} = \tilde{\xi}_{\mu\nu} \, q_{\alpha}q_{\beta} \,
\bar{\psi}\Gamma^{\mu\alpha\rho}\psi\,\bar{\psi}\Gamma^{\nu\beta}_{\;\;\;\;\rho}\psi\, 
\label{cNS2}
\ee
The RR coupling is
\be \Psi^{RR}_{(2)} =\sum_{k~even} \frac 1{k!} C_{(k)}^{i_1 ... i_k} 
\mbox{Tr}_S (\gamma^{i_1 ... i_k}\gamma^{j_1j_2}\gamma^{j_3j_4}M)
q_{l_1}q_{l_2}\omega_{j_1...j_4}^{l_1l_2}  \label{R4} \ee
Using again the gauge condition $q^{i_1} C_{(k)}^{i_1 ... i_k} =0$ and after some
manipulations, similar to those so far performed,
it is not difficult to put eq.(\ref{R4}) into the covariant form
\be
\Psi^{RR}_{(2)} =\sum_{k~even} \frac 1{k!} 
C^{(k)}_{\mu_1 ... \mu_k} \mbox{Tr}_S (\Gamma^{\mu_1 ... \mu_k}
\Gamma_{\nu_1\nu_2} \Gamma_{\nu_3\nu_4}{\cal M})\,q_{\alpha}
q_{\beta}(\bar{\psi}\Gamma^{\nu_1\nu_2\alpha}\psi\,
\bar{\psi}\Gamma^{\nu_3\nu_4\beta}\psi)
\label{cR2}
\ee
Following the same procedure, it is possible to write down all the 
other terms.

\section{Spin effects for the p-p system}

Let us consider now spin interactions between two parallel 
slowly moving Dp-branes with impact parameter $\vec b$. 
Recall that we identify the ``time'' with $X^1$; accordingly, the boundary state 
of a brane moving with velocity $v^i$ is obtained from the static one by applying 
the boost operator $e^{iv_iJ^{1i}}$ \cite{billo}, where
$$ 
J^{1i}=\oint_{\tau=0}\!d\sigma\left(X^{[1}\partial_{\sigma}X^{i]}
-\frac{i}{4}\bar{S}\rho^1\,\gamma^{1i}S\right) 
$$
choosing to boost the boundary state defined in $\tau=0$. Since, on the boundary,
$iM^{(s)}\tilde{S}=S$, the vertex operator $V_B\equiv iv_iJ^{1i}$ can be 
written as 
\be 
V_B=v_i\oint_{\tau=0}\!d\sigma\left(X^{[1}\partial_{\sigma}X^{i]}
+\frac{1}{2}S\,\gamma^{1i}S\right)
\label{vb} 
\ee
where now $S$ is just the left-moving part of the world-sheet spinor. The same
operator (\ref{vb}) could also have been derived from that of photons in Type I 
theory with a constant field strength background after a T-duality transformation
\cite{bach}.

As discussed in the introduction, leading orders in the expansion in powers
of $v$ can be read from correlations including the appropriate power
of $V_B$ insertions in the static brane-brane potential eq.(\ref{cyli}).
Before going on, it is important to point out that in computing leading orders of 
velocity-dependent potentials through correlation functions, we can actually directly 
extract potentials from the corresponding phase-shifts by simply dropping the overall
time factor; this can be easily understood remembering that the bosonic 
coordinates along the velocity direction are not twisted and, according to 
eq.(\ref{conf}), the resulting zero mode integration turns then the phase shift into 
the potential, evaluated at the space-time $T=0$, i.e. the time when the two D-branes 
are passing one each other at distance $\vec{b}$.
 
Given these preliminaries, we can evaluate correlation functions
involving in general $n$ $V_B$'s and $m$ $V_\eta$'s.
The corresponding amplitudes are given by
\begin{equation}
{\cal A}_{n,m}=\frac{1}{16n!(2m)!}\left(\begin{array}{c} 
2m \\ m \end{array}\right)
\int_0^\infty \!\!dt \, 
\langle B_p,\vec{x}=0|e^{-2\pi t\al p^+(P^--i\partial/\partial x^+)}
(V_B)^n(V_{\eta})^{m}|B_p,\vec{y}=\vec{b} \rangle
\label{cyl}
\end{equation}
where the combinatorial factors come from the expansions of the boost and
supersymmetry operator, eq.(\ref{u}). There is an evident
analogy between eq.(\ref{cyl}) and four-point 1-loop amplitudes of massless states, 
in Type I string theory in the G-S formalism. In particular, the zero mode trace
is vanishing unless all the eight zero modes $S_0$ are inserted \cite{gsw}, i.e.
$$
\langle B_0| R_0^N|B_0\rangle = \mbox{Tr}_{V}[R_0^N]-
\mbox{Tr}_{S}[R_0^N]=0 \;,\;\; \mbox{for}\ \  N<4 
$$
where the trace and matrix multiplication in both terms are over the vectorial and
spinorial indices. Since $V_B$, as well as $V_\eta$, provides at most two of them,
a total of $n+m \geq 4$ vertex insertions is needed in order to get a non-zero result.
The first non-vanishing trace is
\bea
t^{i_1...i_8} &\equiv& \mbox{Tr}_{S_0}\,R_0^{i_1i_2}R_0^{i_3i_4}R_0^{i_5i_6}R_0^{i_7i_8} 
\nn \\ &=& - \frac 12 \epsilon^{i_1 ... i_8} - \frac 12 
\left[\delta^{i_1 i_4}\delta^{i_2 i_3}\delta^{i_5 i_8}\delta^{i_6 i_7}
+ \mbox{perm.}\right] \nn \\ 
&\;& + \frac 12 \left[\delta^{i_2i_3} \delta^{i_4i_5}
\delta^{i_6i_7}\delta^{i_8i_1} + \mbox{perm.} \right]
\label{t}
\eea
where ``perm.'' means permutations of the pairs $(i_{2n-1}i_{2n})$ plus antisymmetrization 
within all the pairs.

We will consider in the following the special case $n+m=4$. This corresponds, for a fixed $m$, 
to the leading order in the velocity of the associated spin potential. The interest
of this case lies in the fact that, being determined only by the fermionic zero mode part
of both vertex insertions $V_B$ and $V_\eta$, massive string contributions precisely cancel, 
exactly as in eq.(\ref{static}). These amplitudes are therefore {\it scale invariant}, in the 
sense that their dependence on the distance $\vec b$ is exact, keeping the same functional 
form at any finite distance.
In the following, expressions similar to eq.(\ref{cyl}) will be denoted simply by
${\cal A}_n \equiv \langle V_B^n\,V_{\eta}^{4-n}\rangle$, in order to light the notation.
We wrote in eq.(\ref{cyl}) all the supercharges applied to the same boundary state; 
being fixed simply by a zero modes analysis, the computation will not depend on the
choice of the boundary, whereas the physical interpretation as polarization effects 
will be different. 
The polarizations of the two D-branes are indeed given by the supersymmetric
parameters $\eta_1^a,\eta_1^{\dot{a}}$ and $\eta_2^a,\eta_2^{\dot{a}}$ associated
to the two boundaries, as shown in figure 3.

\vskip 20pt

\input epsf
\epsfsize=50pt
\centerline{\epsffile{2p.eps}}
\vskip 10pt
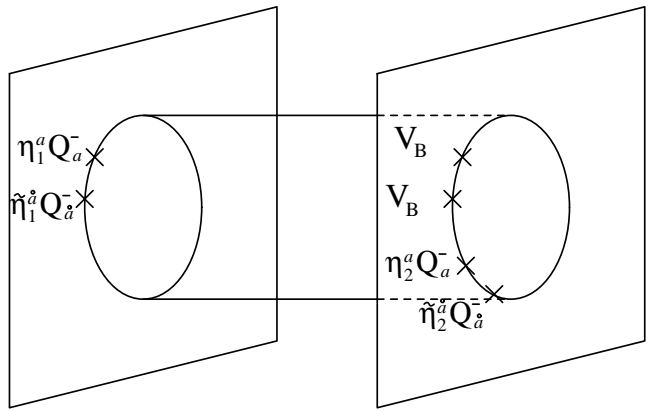
\centerline{{\bf Fig. 3}: example of a spin-dependent term at $v^2$ order, $m=n=2$.}    

\vskip 10pt

Let us start by inserting only the bosonic vertex operators $V_B$, that means to 
consider the universal $v^4$ potential ${\cal A}_4=\langle V_B^4 \rangle$.
From the above analysis, it follows that a non-vanishing result is 
obtained only when we pick up the fermionic part of the operator (\ref{vb}) and in
particular the zero modes for each operator $S$. 
The computation is straightforward and the result is
\be 
{\cal A}_4=\frac{V_{p+1}}{64}\,T_p^2\,|v|^4\,G_{9-p}(\vec{b}) 
\label{a4} 
\ee
As well known, possible contributions to a static force or to $v^2$-potentials
are absent due to a compensation between the gravitational and dilatonic fields 
(attractive) and the RR $A_{p+1}$ field, that for two Dp-branes is repulsive, of 
course. In this formalism, it is immediately clear that supersymmetry implies a 
contribution starting only like $v^4$. 

The first spin effect is given by 
${\cal A}_3= \langle V_B^3\,V_{\eta} \rangle$; going through the same steps and 
after some simple algebra, one obtains
\beqa  
{\cal A}_3&=&\frac{V_{p+1}}{8}|v|^2\,(4\pi^2\al)^{4-p}
\int_0^\infty dt \int\frac{d^{9-p}q}{(2\pi)^{9-p}}\,
e^{i \vec{q}\cdot (\vec{x}-\vec{y})-2\pi t\al q^2}
q_j\omega_{1i}^j(\eta)\,v^i \nn \\
&=&-i\frac{V_{p+1}}{16}T_p^2\,|v|^2v^i\,\tilde{\eta}\gamma^{1ij}\eta \,
\partial_j G_{9-p}(\vec{b}) \label{a3} 
\eeqa\noindent
that represents a spin-orbit like coupling between branes. From eqs.(\ref{NS}) 
and (\ref{R}) we can derive the NSNS and RR polarizations of the exchanged states,
responsible of these
non-minimal couplings. In order to perform the analytic continuation of 
eq.(\ref{a3}) to Minkowskian coordinates, it is convenient to write covariantly the 
term $\tilde{\eta}\gamma^{1ij}\eta$, whose $SO(1,9)$ expression is
$\bar\psi\Gamma^{1\mu\nu}\psi$.
Performing the rotation we obtain $i\bar{\psi}\Gamma^{0\mu\nu}\psi$, and sending 
$v^i\rightarrow iv^i$ leads finally to
\be  {\cal A}_3^{Mink.}=-\frac{V_{p+1}}{32}T_p^2\,|v|^2v_{i}\,
\partial_{j} G_{9-p}(\vec{b})\, J^{0ij}
\label{a3m}  \ee
where $i,j=1,...,9$ and $J^{\mu\nu\rho} \equiv i\bar{\psi} \Gamma^{\mu\nu\rho}\psi$
The next spin effect is ${\cal A}_2= \langle V_B^2\,V_{\eta}^2 \rangle$; in this 
case we have to distinguish two possible configurations, depending to which  
boundary state we apply the supercharges:
$$ 
{\cal A}_2^{(1)}=\langle V_B^2\,V_{\eta}^2 \rangle; \ \ \ 
{\cal A}_2^{(2)}=\langle V_{\eta_1}\,V_B^2\,V_{\eta_2} \rangle  
$$
These two contributions can be written as
\beqa  
{\cal A}_2^{(1)}&=&\frac{V_{p+1}}{32}T_p^2\,\omega_{i_1...i_4}^{ij}(\eta)
\,t^{i_1...i_41k1l}\,v_k v_l\partial_i\partial_j G_{9-p}(\vec{b}) \label{a2} \\
 {\cal A}_2^{(2)}&=&\frac{V_{p+1}}{32}T_p^2\,\omega_{[i_1i_2}^{i}(\eta_1) \label{a2b}
\omega_{i_3i_4]}^{j}(\eta_2)
\,t^{i_1...i_41k1l}\,v_k v_l\partial_i\partial_j G_{9-p}(\vec{b}) 
\eeqa
Noting that
\be
t^{i_1...i_41k1l}\,v_k v_l=v^2\,(4\delta^{1i_2}\delta^{1i_4}\delta^{i_1i_3}-
\delta^{i_1i_3}\delta^{i_2i_4})-4v^{i_1}v^{i_4}\delta^{i_2i_3}
\ee
and working out the spinor algebra, it can be shown that eq.(\ref{a2}) reconstructs the
covariant amplitude, that after analytic continuation, takes the following form:
\be
{\cal A}_2^{(1)}=\frac{V_{p+1}}{768}T_p^2\,v^2
(2 J^{m 0q} J^{n}_{\;\; 0q} - J^{m p q} J^{n}_{\;\; p q}
+ 4 J^{m \rho}_{\quad \; i} J^{n}_{\;\; \rho j} \hat v^i \hat v^j)
\partial_m \partial_n G_{9-p}(\vec{b}) \label{a2m}
\ee
Latin letters $i,j,k,...$ label $SO(9)$ indices running from 1 to 9, in contrast to 
$SO(1,9)$ indices, denoted with Greek letters.
In the same way, one can reconstruct the explicit covariant form of eq.(\ref{a2b})
and all the remaining spin effects that will follow.
We do not report the explicit relations for all the cases, being quite lengthy, 
as well as the analytic continuation. The remaining spin effects are
\beqa  
{\cal A}_1^{(1)}&=&\langle V_B\,V_{\eta}^3 \rangle=
\frac{V_{p+1}}{144}T_p^2\,\omega_{i_1...i_6}^{ijk}(\eta)
\,t^{i_1...i_61l}\, v_l\partial_i\partial_j\partial_k G_{9-p}(\vec{b}) 
\label{a1} \\
{\cal A}_1^{(2)}&=&\langle V_{\eta_1}\,V_B\,V_{\eta_2}^2 \rangle=
\frac{V_{p+1}}{144}T_p^2\,\omega_{i_1i_2}^{i}(\eta_1)
\omega_{i_3...i_6}^{jk}(\eta_2)
\,t^{i_1...i_61l}\, v_l\partial_i\partial_j\partial_k G_{9-p}(\vec{b}) \nn 
\eeqa
and the static force
\beqa  
{\cal A}_0^{(1)}&=&\langle V_{\eta}^4 \rangle=
\frac{V_{p+1}}{4(4!)^2}T_p^2\,\omega_{i_1...i_8}^{ijkl}(\eta)
\,t^{i_1...i_8}\, \partial_i\partial_j\partial_k\partial_l G_{9-p}(\vec{b})
\nn \\
{\cal A}_0^{(2)}&=&\langle V_{\eta_1}\,V_{\eta_2}^3 \rangle=
\frac{V_{p+1}}{4(4!)^2}T_p^2\,\omega_{i_1i_2}^{i}(\eta_1)
\omega_{i_3...i_8}^{jkl}(\eta_2)
\,t^{i_1...i_8}\, \partial_i\partial_j\partial_k\partial_l G_{9-p}(\vec{b})
\nonumber \\
{\cal A}_0^{(3)}&=&\langle V_{\eta_1}^2\,V_{\eta_2}^2 \rangle=
\frac{V_{p+1}}{4(4!)^2}T_p^2\,\omega_{i_1...i_4}^{ij}(\eta_1)
\omega_{i_5...i_8}^{kl}(\eta_2)
\,t^{i_1...i_8}\, \partial_i\partial_j\partial_k\partial_l G_{9-p}(\vec{b})
\label{a0}\nn 
\eeqa
In all these cases the one-point functions considered in last section allows to
see which are, in each configuration, the massless string excitations responsible of
all these polarization effects.

\section{Spin effects for the p-p+4 system}

Let us now consider spin potentials for parallel p-p+4 brane
configurations. Like more general p-q systems with mixed Neumann-Dirichlet
boundary conditions in four directions, these BPS configurations preserves
1/4 of the initial supersymmetries, rather than the 1/2 of the p-p
system. This residual supersymmetry protects as before the leading order 
term in the velocity from higher massive modes contributions;
however, unlike the p-p system, the reduced amount of 
supersymmetry allows now a non-trivial metric in the
Dp moduli space. In particular the D0-D4 system, studied
in \cite{dkps}, was proposed in \cite{douglas}
as a matrix description for the scattering of an eleven
dimensional supergraviton off the background of a longitudinal fivebrane. 
The aim of this section is to study the leading spin dependence of the potential felt
by slowly moving p-branes in the p+4 background.
The relevant zero mode traces are in this case of the form
\be
\langle B_p|{\cal O}|B_{p+4} \rangle =\mbox{Tr}_{S_0}[{\cal O}N] 
\equiv \mbox{Tr}_V [{\cal O}N] - \mbox{Tr}_S [{\cal O}N]
\label{bpbp4}
\ee  
where ${\cal O}$ is a product of $R_0^{ij}$ arising from 
the zero mode part of the $V_B$ and $V_\eta$ vertex insertions and
\bea
N^{ij} &\equiv& (M_p^T M_{p+4})^{ij} = \pmatrix {
I_{p+1}& 0 & 0 
\cr 0 & -I_4 & 0 
\cr 0 & 0 & I_{3-p} \cr} \nn \\ 
N_{\dot a \dot b} &\equiv& (M_p^T M_{p+4})_{\dot a \dot b} 
= (\gamma^{p+2} ... \gamma^{p+5})_{\dot a \dot b}
\label{N}
\eea
By simple inspection of eq.(\ref{bpbp4}), using the matrices (\ref{N}) and the
representation of the operators (\ref{R0}), we get vanishing 
traces for ${\cal O}=1,R_0^{ij}$. The first non trivial result is 
\bea 
t^{i_1...i_4} &\equiv& \mbox{Tr}_{S_0} \,R_0^{i_1i_2}R_0^{i_3i_4} \nn \\
&=&2 \, \epsilon^{i_1... i_4 p+2... p+5} \nn \\
&\;& +2\left(\delta^{i_1p+2}\delta^{i_2 p+3}\delta^{i_3p+4}\delta^{i_4p+5} 
+N^{i_2 i_4}\delta^{i_1i_3}+ \mbox{perm.} \right)
\label{tprime}
\eea
where by ``perm.'' we mean as before an antisymmetrization within each 
pair $(i_1,i_2)$, $(i_3, i_4)$ and symmetrization under the exchange
of each of these pairs.  

The relevant amplitudes describing leading spin-effects are then  
$$
{\cal A}_n=\frac{1}{16n!(4-2n)!}\left(\begin{array}{c} 4-2n \\ 2-n \end{array}\right)
\int_0^\infty \!\!dt \, 
\langle B_p,\vec{x}=0|e^{-2\pi t\al p^+(P^--i\partial/\partial x^+)}
(V_B)^n(V_{\eta})^{2-n}|B_{p+4},\vec{y}=\vec{b} \rangle
$$
where the total number of vertex insertions now is two providing the four
zero modes required to get the first non-trivial result from eq.(\ref{tprime}).
The rest of the computation follow the lines of last section. We are left 
with the universal term  
\be 
{\cal A}_2=\frac{V_{p+1}}{4}\,T_p T_{p+4}\,|v|^2\,G_{5-p}(\vec{b})
\label{b2} 
\ee
and the leading spin potentials
\beqa  {\cal A}_1&=&\frac{V_{p+1}}{4}T_p T_{p+4}\,
\omega^i_{i_1 i_2}(\eta)\, t^{1ji_1 i_2}\,v_{j}\,\partial_{i} G_{5-p}(\vec{b})\nonumber\\  
{\cal A}_0^{(1)}&=&\frac{V_{p+1}}{16} T_p T_{p+4}\,\omega_{i_1...i_4}^{ij}(\eta)
\,t^{i_1... i_4}\,\partial_i\partial_j G_{5-p}(\vec{b}) \nonumber \\
{\cal A}_0^{(2)}&=&\frac{V_{p+1}}{16} T_p T_{p+4}\,\omega_{i_1i_2}^{i}(\eta_1)
\omega_{i_3i_4}^{j}(\eta_2)
\,t^{i_1...i_4}\,\partial_i\partial_j G_{5-p}(\vec{b})
\label{b0} 
\eeqa
The appearence of $T_{p+4}$ and $G_{5-p}$ instead of the 
$T_p$ and $G_{9-p}$ for the p-p system is due to the lack of four
Dirichlet-Dirichlet transfered momentum integrations.   

We recall that eqs.(\ref{b2}) and (\ref{b0}) are exact
to any order in the brane separation $\vec b$.
Of course this is a peculiar property only of these leading order terms and of
the supersymmetric p-p, p-p+4 configurations. Higher order terms or
more general brane configurations will involve contributions from
the oscillator part of the vertices (\ref{vb}), (\ref{veta}) described by
modular functions with non-trivial transformation properties which in
general distinguish the large and short distance behaviors. 
We should say however that this 
property is shared by an amount of other interesting brane systems. 
In a final discussion we will go on through many of these examples showing how
supersymmetry is enough to ensure the scale invariance of all their 
relevant leading interactions.

\section{Field theory interpretation and D0-brane \\ gyromagnetic ratio}

In the present section we discuss the field theory interpretation of our results. We will
show in particular that the knowledge of all the one-point functions of the massless 
fields of Type IIA/B supergravity allows to infer the complete and generic asymptotic
form of the corresponding p-brane solution. Moreover, the spin-effects in scattering 
amplitudes that we have computed in section 4 and the supersymmetric 
cancellation of some of their leading orders proves to constitute an 
extremely efficient way to fix unambiguously the various coefficients entering the 
solution, and in particular the relative strength of the NSNS attraction and the RR 
repulsion (the fact that normalizations are better encoded in scattering amplitude than 
in one-point functions, especially through the vanishing of leading order, was already 
appreciated in Polchinski's computation of the Dp-brane charge \cite{pol}). As we will
see, this approach yields a powerful technique to extract informations
about a generic component of the p-brane multiplet. The analogous
computation in supergravity would consist in performing supersymmetry transformations
to the usual p-brane solution, to determine all the spinning superpartners;
this requires looking up to eight variations, a program that, as can be appreciated from 
previous works \cite{aich,dlrbig,dlr}, is out of reach within the component fields 
formalism.

Collecting the covariant one-point functions (\ref{cov}), (\ref{cNS1}), (\ref{cR1}), (\ref{cNS2}) 
and (\ref{cR2}), for up to four supercharge insertions, the NSNS 
and RR asymptotic fields for a generic component of the Dp-brane multiplet can be written as 
a multipole expansion in momentum space:
\bea
\xi^{\mu\nu} &=& \kappa^2 
\left[A_0 M^{\mu\nu} + A_1 J^{\mu \sigma \alpha} M^\nu_\sigma q_\alpha 
+ A_2 J^{\mu \alpha \rho} J^{\sigma \beta}_{\;\;\; \rho}
M^\nu_\sigma q_\alpha q_\beta + ... \right] \label{fieldNS} \\
C_{(k)}^{\mu_1 ... \mu_k} &=& \frac {\kappa^2}{k!} \left[
B_0 \mbox{Tr}_S [\Gamma^{\mu_1 ... \mu_k} {\cal M}] +
B_1 \mbox{Tr}_S [\Gamma^{\mu_1 ... \mu_k} \Gamma_{\nu_1 \nu_2} {\cal M}]
J^{\nu_1 \nu_2 \alpha} q_\alpha \right. \nn \\ 
&\;& \quad \;\; + \left. B_2 
\mbox{Tr}_S [\Gamma^{\mu_1 ... \mu_k} \Gamma_{\nu_1 \nu_2} \Gamma_{\nu_3 \nu_4} {\cal M}] 
J^{\nu_1 \nu_2 \alpha} J^{\nu_3 \nu_4 \beta} q_\alpha q_\beta + ... \right] 
\label{fieldR}
\eea
We have restored the ten-dimensional Plank constant $\kappa^2$ for convenience.  
Dots stand for six and eight supercharge insertions, corresponding to three and four powers of 
momentum, that we shall not consider. The constants $A_i,B_i$ could in 
principle be fixed by correctly normalizing the one-point functions reported in
section 3; however, 
this is awkward, and since any final conclusion will eventually 
depend in a crucial way on these constants, we will take advantage of our 
results for the scattering amplitude to fix them unambiguously.

From now on we specialize to the D0-brane, for which $M^0_0 = -1$, $M^i_j = 
\delta^i_j$ and ${\cal M} = \Gamma^0$; the other cases can be treated in the same way.
Recall that in the NSNS sector, a 
generic field $\xi_{\mu\nu}$ is decomposed into trace, symmetric and 
antisymmetric parts $\phi$, $h_{\mu\nu}$ and $b_{\mu\nu}$ as
$$
\epsilon_{\mu\nu}^{(\phi)} = \frac 14 (\eta_{\mu\nu} - q_\mu l_\nu - q_\nu l_\mu) 
\;,\;\; \epsilon_{\mu\nu}^{(h)} = \xi_{(\mu\nu)} \;,\;\; 
\epsilon_{\mu\nu}^{(b)} = \xi_{[\mu\nu]} $$ 
where $l^\mu$ is a vector satifying $q \cdot l = 1$, $l^2=0$. 
The asymptotic fields in the NSNS sector 
are then found to be
\bea
&&\phi = \frac 32 \kappa^2 M G_9(r) + \frac 14 \kappa^2 C 
J^{mpq} J^{n}_{\;\; pq} \partial_m \partial_n G_9(r) + ... \nn \\
&&\left\{
\begin{array}{l}
\displaystyle{h_{00} = \kappa^2 M G_9(r) + \kappa^2 C 
J^{m0q} J^{n}_{\;\; 0q} \partial_m \partial_n G_9(r) + ...} \nn \medskip\ \\
\displaystyle{h_{ij} = \delta_{ij} \kappa^2 M G_9(r) + \kappa^2 C 
J^{m \;\; \rho}_{\;\;\; i} J^{n}_{\;\; j \rho} 
\partial_m \partial_n G_9(r) + ...} \nn \medskip\ \\
\displaystyle{h_{0i} = 2 \kappa^2 A J_{0i}^{\;\;\; m} \partial_m G_9(r) + ...} \nn
\end{array}
\right. \\ && \left\{
\begin{array}{l}
b_{ij} = \kappa^2 A J_{ij}^{\;\;\; m} \partial_m G_9(r) + ... \medskip\ \\
b_{0i} = 2 \kappa^2 C J^{m}_{\;\; 0q} J^{n \;\; q}_{\;\;\; i} 
\partial_m \partial_n G_9(r) + ...
\label{fNS}
\end{array}
\right.
\eea
whereas eq.(\ref{fieldR}) in the RR sector become
\bea
&&\left\{
\begin{array}{l}
C_0 = 2 \kappa^2 Q G_9(r) + \kappa^2 D 
J^{m \rho \tau} J^{n}_{\;\; \rho \tau} \partial_m \partial_n G_9(r) + ... \nn \medskip\ \\
C_i = 2 \kappa^2 B J_{0i}^{\;\;\; m} \partial_m G_9(r) + ... \nn 
\end{array}
\right. \\ && \left\{
\begin{array}{l}
C_{0ij} = \kappa^2 B J_{ij}^{\;\;\; m} \partial_m G_9(r) + ... \medskip\ \\
C_{ijk} = 2 \kappa^2 D J^{m}_{\;\; 0[i} J^{n}_{\;\; jk]} 
\partial_m \partial_n G_9(r) + ...
\end{array}
\right.
\label{fR}
\eea
The constants $A_i,B_i$ have been redefined and called $M,A,B,Q,C,D$ for later convenience,
and again, dots stand for higher derivative terms associated to further supercharge insertions.

Comparing eqs.(\ref{fNS}) and (\ref{fR}) with the usual 0-brane solution \cite{hs}
and the general result valid in D dimensions derived in \cite{mp}, we conclude that 
$M$ is the mass and $Q$ the electric charge, whereas $2 A J_{0ij} = J_{ij}$
is the angular momentum and $B J_{0ij} = \mu_{ij}$ the magnetic moment, so that the 
gyromagnetic ratio, defined by the relation $\mu^{ij}=(gQ)/(2M)J^{ij}$, is given by
$g = (M B)/(Q A)$. Also, the electric and gravitational dipole moments vanish, since 
they would correspond to one-derivative terms in $C_0$ and $h_{00},h_{ij}$ respectively.
On the contrary, the presence of non-vanishing two-derivative terms in eqs.(\ref{fNS}) shows the presence of a gravitational quadrupole moment for 
D-particles. Similarly, there are two-derivative terms also in eqs.(\ref{fR})
corresponding to gauge quadrupole effects. Actually, the one in $C_0$ vanishes
due to a Fiertz identity, showing that D-particles have a vanishing electric
quadrupole moment with respect to the RR one-form.
However, analogously to the gyromagnetic ratio $g$, we can define the ratio of the gauge
and quadrupole moments by $\tilde g = 4 (M D)/(Q C)$.   

It is now straightforward to show how the semiclassical analysis of the phase-shift between
two of these configurations can be used to determine in a simple way the value of the 
gyromagnetic ratio $g$ and its quadrupole analogue $\tilde g$ associated to 
D0-branes. According to \cite{hioy,dlr}, massive Kaluza-Klein states present a
common value $g=1$, contrarily to the usual and ``natural'' \cite{wei,fpt,jack} value 
$g=2$ shared by all the known elementary particles (neglecting radiative corrections, 
of course). This particular signature of Kaluza-Klein states can be useful to 
establish the 11-dimensional nature of D0-branes, implying $g=1$. This consistency check 
has been recently performed in \cite{dlr} considering D0-branes as extended extremal 
0-brane solution of IIA supergravity. We present an alternative and independent argument 
that relies on the ``stringy'' nature of D0-branes as points on which open strings can end; 
in particular, we show that $g=1$ is the only possible value compatible with the cancellation 
of the linear term in velocity in the first spin effect, eq.(\ref{a3m})
\footnote{Actually the fact that the value of $g$ is related to the 
cancellation of the leading term linear in $v$ was implicit in ref.\cite{noi,kra}, even 
if its significance was not completely recognized.}. Similarly, we will show that our 
stringy analysis predicts for the quadrupole analog the value $\tilde g = 1$ from the 
cancellation of the static contribution to the second spin effect, eq.(\ref{a2}). 

Consider first the scattering of a scalar 0-brane, taken as a probe, off a spinning
0-brane, acting as source. The effective
action for the probe is (in the string frame)
\be
{\cal S} = - M\int \!d\tau \,e^{- \phi}\sqrt{-g_{\mu\nu} \dot X^\mu \dot X^\nu} - 
Q\int \!d\tau C_\mu \dot X^\mu
\label{1ps} 
\ee 
For a trajectory with constant velocity $v=\tanh \pi \epsilon$, we can choose
$X^0(\tau)=\tau\,\cosh\pi\epsilon$, $X^i(\tau)=\tau\,\hat{v}^i\sinh\pi\epsilon$. 
Expanding for small velocities and weak fields
($\kappa \rightarrow 0$), one finds, dropping a constant term, 
${\cal S} = \int \! d\tau \sum_{n\ge 0} v^n {\cal L}_n$ with
\bea
&&{\cal L}_0 = M \phi + \frac 12 M h_{00} - Q C_0 \nn \\
&&{\cal L}_1 = M h_{0i} \hat v^i - Q C_i  \hat v^i \;,\;\;
{\cal L}_2 = \frac 12 M (h_{00} + h_{ij} \hat v^i \hat v^j) 
- \frac 12 Q C_0 \nn \\
&&{\cal L}_3 =  M h_{0i} \hat v^i - \frac 12 Q C_i \hat v^i \;,\;\;
{\cal L}_4 = \frac 12 M (h_{00} + h_{ij} \hat v^i \hat v^j) - \frac 38 Q C_0
\eea
We know from the amplitudes computed in section 4 that the leading non-vanishing contributions
to the scattering amplitude behave like $v^n/r^{7+n}$, all lower orders in velocity 
cancelling by supersymmetry. Substituing the relevant asymptotic fields of the 
spinning 0-brane from eqs.(\ref{fNS}) and (\ref{fR}), one then finds the following 
conditions:
\bea
&& \left. {\cal L}_0 \right|_{G} = 0 \; \Rightarrow \; M = Q \;,\;\; 
\left. {\cal L}_0 \right|_{\partial^2 G} = 0 \; \Rightarrow \; M C  = 4 Q D \nn \\ 
&& \left. {\cal L}_1 \right|_{\partial G} = 0 \; \Rightarrow \; M A = Q B \label{coef} \\
&& \left. {\cal L}_2 \right|_{G} = 0 \; \Rightarrow \; M = Q \nn
\eea 
Altogether, this yields
\be
Q = M \;,\;\; g = 1 \;,\;\; \tilde g = 1 
\ee
One can now use these results to write down 
the structure of the leading non-vanishing contributions to the scattering amplitude.
Actually, thanks to the results obtained in section four we can write down 
the amplitude with all the coefficients fixed.  Up to an overall factor, we get
\bea
{\cal A} = &&  \kappa^2 M^2 v^4 G_9(r) 
+ 2 \kappa^2 M v^3 J_{0i}^{\;\;\; m} \hat v^i \partial_m G_9(r)
\nn \\ && + \frac 1{12} \kappa^2 M v^2 (2 J^{m 0q} J^{n}_{\;\; 0q} 
- J^{m p q} J^{n}_{\;\; p q}
+ 4 J^{m \rho}_{\quad \; i} J^{n}_{\;\; \rho j} \hat v^i \hat v^j)
\partial_m \partial_n G_9(r)\, + ...
\label{boh}
\eea
The matching of the tensor structure of the $v^3$ and $v^2$ terms with expressions 
(\ref{a3m}) and (especially) (\ref{a2m}) is a non-trivial check of the consistency of the 
whole picture.

A comment is in order on how our boundary state formalism for describing higher 
spin Dp-branes is related to the supergravity description, where p-branes appear as 
solitonic solutions. As already said, the asymptotic fields found by applying the procedure of 
this section correspond to supergravity solutions obtained by taking supersymmetric 
variations of the usual scalar ones. This has been partially done in \cite{dlr} for the 
D0-brane solution, where the second supersymmetry variation was used to compute the angular 
momentum dependence of $h_{\mu\nu}$ and $C_\mu$. Using the same strategy, we have similarly
checked that the angular momentum contributions to the higher forms 
$b_{\mu\nu}$ and $C_{\mu\nu\rho}$ 
(which have not been considered in \cite{dlr}) correctly reproduce those in 
eqs.(\ref{fNS}) and (\ref{fR}). 
We have also checked that the fourth supersymmetry variation reproduces
all the two-derivative terms we find, but it is unrealistic to compute and trust the coefficient
because of the increasing complexity of the involved expressions.  

Finally, another interesting outcome of the knowledge of the asymptotic fields (\ref{fNS}), (\ref{fR}) 
is the possibility to derive the supersymmetric completion of the 0-brane world-line
effective action (\ref{1ps}) {\it in an arbitrary Type IIA background}, at least for
weak fields. For example, it is not dificult to verify that, in much the same way as the
part of the asymptotic fields going like $1/r^7$ can be derived from the linearization
of the action (\ref{1ps}), the part of the fields going like $1/r^8$ can be derived
from the following non-minimal couplings
\be
\label{s2Q} 
{\cal S}_{2Q} =  \int \! d \tau \, \left[ -\partial_i h_{0j} J^{0ij}
+ \frac 14 \partial_i b_{jk} J^{ijk} +  \partial_i C_j J^{0ij}
- \frac 14\partial_i C_{0jk} J^{ijk} \right]
\ee
The coefficients have been further checked by computing the static force contribution
of order $1/r^9$ between two spinning 0-branes, that vanishes as expected.

Notice that the covariant form of eq.(\ref{s2Q}) should be obtained by replacing each $0$ 
index by a contraction with the momentum $\dot X^\mu$; in such a way, the fields generated by a 
moving 0-brane are given by the boost of those produced by a static one. One obtains  
\be
{\cal S}_{2Q} = \int \!d\tau \, \left[\Gamma^\rho_{\;\; \sigma \mu} 
\dot X^\mu \dot X^\nu J_{\rho \;\; \nu}^{\;\; \sigma} + 
\frac 1{24} H_{\mu \nu \rho} J^{\mu \nu \rho} 
+ \frac g2 F_{\mu \nu} \dot X^\rho J^{\mu \nu}_{\quad \rho} 
+ \frac g {72} F_{\mu \nu \rho \sigma} \dot X^\mu J^{\nu \rho \sigma} 
\right] 
\label{acov}
\ee
where $F_{\mu \nu}$, $F_{\mu \nu \rho \sigma}$ and $H_{\mu \nu \rho}$ are 
the field-strengths of $C_\mu$, $C_{\mu \nu \rho}$ and $b_{\mu\nu}$. We recognize a two-fermion 
gravitational term showing up through a coupling to the Christoffel connection
$\Gamma^\mu_{\;\; \alpha \beta}$, coming from the (linearized) spin-connection
entering the spinor covariant derivative, and similar non-minimal couplings to the 
gauge field curvatures. At next order, a four-fermion term manisfesting itself through 
a coupling to the Riemann tensor is expected, among further non-minimal couplings.
The coupling to the higher RR forms and the NSNS antisymmetric 
tensor seems to occur through more complicated terms which correctly 
disappear in the absence of fermionic background. From the eleven-dimensional point 
of view, eq.(\ref{acov}) is the Kaluza-Klein reduction  
(keeping $J$ to be ten-dimensional) of an action containing only the first and
last terms for the eleven-dimensional metric and three form. In order to work, this requires 
$g=1$, as it is.

\section{Final remarks and conclusions}

We studied in the present paper interactions between brane configurations associated
to two parallel p-p and p-p+4 branes, using the Green-Schwarz boundary state
formalism. We found explicit expressions for spin-dependent interactions between 
moving branes. Our general strategy can be summarized as follows: instead of
considering the full configuration of moving branes, where supersymmetry is broken,
we perturbed through appropriate vertex operators the supersymmetric vacuum associated 
to the static p-p (p-p+4) system,
allowing in this way to easily derive important results 
on the structure of the exact (in powers of $\alpha^\prime$)  
leading spin interactions in a velocity expansion. 
The cylinder computation of these terms collapses 
to its zero mode contribution,
supporting an equivalent description 
in terms of either the open (matrix model) or  
closed (supergravity) massless degrees of freedom. 
However the analysis performed on the relevant cylinder correlations is
quite general and can be easily extended to several other previously studied 
brane configurations (see for example \cite{lif,AB,lm,lifs,tse,pierre})
where a similar long-short 
distance matching of their leading
interactions were observed. They fall in general into two main groups:
supersymmetric brane configurations, which include besides the examples studied
above, the p-p+8 systems, bound states between 
p-p+2, p-p+2-p+4, p-p+4, ... D-branes, and any S or T-dual combinations
of these systems; and brane configurations which are supersymmetric only 
in a certain limit of their moduli space. Bound states can be considered in
general as fluxes for the gauge field living on the boundary of the biggest
D-brane, modifying therefore its boundary conditions. The corresponding 
light-cone boundary state for a generic condensate was constructed  
in \cite{gregut1}. The cylinder amplitude defined by two 
of these boundary states,
in the case that some of the supersymmetries are preserved (take for example
two indentical p-p+2 bound states or the S-dual analog of two D-strings with 
equal electric fluxes turned on \cite{calkl}) , 
will lead to similar vanishing
traces as in eq.(\ref{bpbp4}), unless $N/4$ velocity insertions soak the 
$N/2$ left zero modes, $N$ being the number of supercharges left unbroken
by the system. Again the spin-dependent dynamics can be studied by 
inserting supercharges on the cylinder, and once more an equivalent 
matrix-supergravity description for the slowly moving regime is garanteed.

The second interesting class of configurations (and less straight) for which an
analysis along the lines of this paper can be followed, 
is inspired from the brane systems studied in 
\cite{lm}, which although
not supersymmetric, become so in a given limit of
the moduli space. In the analysis of these systems one can follow a 
strategy parallel to the one previously applied to the case of 
moving branes. In that case supersymmetry is broken for finite 
velocity $v$, but the existence of a supersymmetric limit $v=0$
allow us to study leading orders by a simple analysis of the zero mode
structure of amplitudes involving the insertion of vertex
operators corresponding to the deformation (in that case $v$)
from the supersymmetric point. 
Similarly, now we look at the neighborhood of a specific
choice of flux for which some supersymmetry is restored. 
The fermionic part of the operators
corresponding to deformations about this supersymmetric point
coincide with the vertex (\ref{vb}), once
we substitute the plane $(1i)$ defining the boost operation 
with the condensate euclidean plane $(mn)$, and therefore the results can
be read directly from the ones quoted above for the moving brane systems.   
We can ilustrate this with the
simple example of a Dp-brane, wrapped around a $T^2$   
with a magnetic flux $f_1=\frac{N}{2\pi R_m R_n}$
turned on. The boundary state for this
specific condensate can be read from the more general one found in
\cite{gregut1} to be defined by eqs.(\ref{mom}) and (\ref{zm}) through the matrices
\begin{eqnarray}
M_{ij}&=& \sqrt{(1+f^2)}\pmatrix {- I_{p-1}& 0&0&0 \cr
0&\cos{\alpha}&-\sin{\alpha}& 0\cr
0&\sin{\alpha}&\cos{\alpha}&0\cr
0&0&0& I_{7-p} \cr}\nonumber\\
M_{\dot{a}\dot{b}}&=&(\gamma^{12}+ f)\gamma^1\cdots\gamma^{p+1}
\label{Mvc} 
\end{eqnarray}
where $\cos{\alpha}\equiv-\frac{1-f^2}{1+f^2}$. Notice 
that eqs.(\ref{Mvc})
reduce, in the large $f$ limit, to the matrices 
(\ref{Mv}),(\ref{Ms}) defining the D(p-2) 
brane, up to an overall
$f$ factor and the missing of two momentum modes corresponding to
the Neumann-Dirichlet directions. 
As we discussed before, we can study the leading 
interactions of this bound state
with a D(p-2) ordinary brane by simply perturbing the system by a
small $c\equiv 1/f$ quantity from the supersymmetric $c=0$ point. 
The universal potential is then defined by correlators
involving insertions of $R_0^{i1}v^i$ and $R_0^{mn}\pi c$
in the D(p-2)-D(p-2) cylinder, and as before we have vanishing
traces unless at least four of these insertions soak the eight 
zero modes, leaving
\be
\langle B_{p-2}|B_p, c, v\rangle=\frac{1}{\pi c}
\frac{V_{p-1}}{32}T_p T_{p-2}
(v^4+2 v^2(\pi c)^2+ (\pi c)^4) G_{7-p}(r).
\label{bound}
\ee      
The $1/c$ factor can be interpreted as the number of D(p-2) branes
arising from the Dp-brane in the $c\rightarrow 0$ limit, 
while the two missing powers in $r$
represents the reduced transverse space to the system, and the 
relative coefficients are fixed by the kinematical
tensor (\ref{t}) \footnote{A flip in the sign of the $v^2$ comes
from the analytic continuation to the Minkowski plane.}.  
Given, as before, by an exact string computation at the relevant
order in the $v$, $c$ expansion, this potential is valid at
any transverse distance $r$ and in particular admits equivalent
Super Yang-Mills and supergravity descriptions. The p=2 
case is the relevant one for the analysis performed in \cite{lm}.
In that reference the authors study the 
graviton-membrane, static membrane-antimembrane and 
orthogonal moving membranes scattering. In each case the infinite 
boost ($N\rightarrow\infty$ or equivalently $c\rightarrow 0$) represents a
point where the 16 supercharges 
are recovered (for $v=c=0$). The leading orders in $v$, $c$ are
given by eq.(\ref{bound}), and the scale invariance of these terms is garanteed
by our previous analysis, and checked explicitly in that reference.   
The case of orthogonal membranes is particular in the sense that
contains two line of deformations $c\equiv c_1+c_2=0$ 
and $c^\prime\equiv c_1-c_2=0$ (this is the case studied in \cite{lifs}), 
$c_1$, $c_2$ being defined by the fluxes in each 
membrane, along which half the supersymmetries of the D0-D0 system are preserved.
Along these lines, the potential starts then with
$v^2$ as for the previously discussed p-p+4 system. The leading scale invariant
interactions are in general given by
\be
\langle c_2,B_{p2}|B_{p1}, c_1, v\rangle=\frac{1}{\pi^2 c_1 c_2}
\frac{1}{32}T_2 T_2
(v^2+(\pi c)^2)(v^2+(\pi c^\prime)^2) G_{5}(r).
\label{bound2}
\ee  
The absence of the static $c^4$ and ${c^\prime}^4$ terms reflects the
surviving of half-supersymmetries along the aforementioned lines.  
In \cite{lifs,tse} an 
exhaustive list of brane configurations was shown to
present again agreement between the one-loop SYM and semi-classical supergravity 
descriptions of their potentials. Once more, homogenous polynomials of
order four in the fluxes and velocities as in (\ref{bound}),(\ref{bound2}) were found;
an iteration of the analysis for 
the above discussed example provides 
a unified understanding of those results. 
We believe that this example can give a flavour of the generality of
the analysis performed here, which extends to any supersymmetric 
(at least in a point of the moduli space)
brane configuration and covers all (to our knowlegde) one-loop
scattering tests of a given matrix description. We should say that scale invariance 
is however stronger than what a correct matrix
description of supergravity interactions really requires. 
In fact higher
loop potentials will not display a simple decoupling of their massive modes as
in the examples studied here and only a matching between the two
open-closed massless truncated computations can be at most expected.  
Understanding from the string point of view the results quoted 
in \cite{bb,bbpt},
where the $v^6$ potential arising at two loops in the super Yang-Mills 
effective action was shown to agree with the corresponding 
long range correction, or performing other higher loop tests of this
correspondence between SYM and supergravity descriptions of 
D-brane interactions, is an essential step
in the completion of the picture.
It should be pointed out, however, that the light-cone formalism we used becomes 
awkward at higher loops, being affected by several disadvantages.

Finally, we applied our results about the spin-dependent D-brane interactions
to the interesting case of the D-particle. The potential defined by a
cylinder ending on two  
spinning D-particles allowed us to derive supersymmetric analogs of the
universal BPS condition $Q=M$ for the rest of the components of the D0
supermultiplet. In particular we computed the gyromagnetic ratio $g=1$
and a ratio of quadrupole moments $\tilde{g}=1$. The value $g=1$
is the consistent value for D0 branes considered as Kaluza-Klein modes of
eleven dimensional supergravity and was reproduced by a corresponding 
supergravity computation in \cite{dlr}. The results presented 
here can be considered as a further check of the identification of the
0-brane solitonic supergravity solution with the D-particle string definition as 
a point-defect where open strings can end.
Moreover, the string analysis provides a
compact and efficient way to obtain generic supergravity asymptotic solutions 
for spinning p-branes. \\ \\   

\begin{center}{\bf ACKNOWLEDGMENTS}\end{center}

J.F.M. thanks E. Gava and K.S. Narain for useful exchange of ideas.
C.A.S. acknowledges C. Bachas and R. Iengo for interesting discussions.
M.S. thanks the International School for Advanced Studies for hospitality. 
This work has been partially supported by a Pioneer Fund of the Nederlandse Organisatie 
voor Wetenschappelijk Onderzoek (NWO) and EEC contract ERBFMRXCT96-0045.

\appendix

\section{Phase-shift computation in the G-S formalism}
\renewcommand{\theequation}{A.\arabic{equation}}
\setcounter{equation}{0}

In this Appendix, we briefly show how the full phase-shift computation,
performed in the covariant formalism in ref. \cite{bach} as an open string 
vacuum amplitude and in ref. \cite{billo} as an overlap of boundary states, 
is reproduced in the Green-Schwarz light-cone formalism.
As troughout all this article, we will use the double analytic continuation 
described in \cite{noi} in order to define a moving brane; we will therefore 
work in Euclidean space-time and only at the end of the calculation we 
continue analytically our result back to Minkowski 
space-time\footnote{Throughtout this appendix we fix ${\al}=1/2$.}. 

\subsection{Closed string channel}

The moving boundary state can be obtained from the static one, 
eqs. (\ref{mom})-(\ref{conf}), by performing a Lorentz transformation with negative
rapidity $\epsilon$. As explained, this will be performed in Euclidean 
space-time by a transverse rotation generated by the operator 
$J^{ij} = J_B^{ij} + J_F^{ij}$, where
\bea
J_B^{ij} &=& x^i p^j - x^j p^i - i \sum_{n=1}^{\infty} \frac 1n \left(
\alpha_{-n}^i \alpha_n^j - \alpha_{-n}^j \alpha_n^i
- \tilde \alpha_{-n}^i \tilde \alpha_n^j + \tilde \alpha_{-n}^j \tilde 
\alpha_n^i \right) \\
J_F^{ij} &=& - \frac i4 \sum_{n=-\infty}^{\infty}
\left(S^a_{-n} \gamma^{ij}_{ab} S^b_n + 
\tilde S^a_{-n} \gamma^{ij}_{a b} \tilde S^b_n
\right)
\eea
Taking the velocity in the $X^8$ direction, we have to compute
$|B_v \rangle = e^{iv J^{18}} |B \rangle$.
The zero mode part of $J_B^{18}$ breaks translational invariance along the
$X^8$ direction and turns eq.(\ref{conf}) into 
\be
|B_v,\vec{x}\rangle=(\sqrt{2} \pi)^{4-p}\int\frac{d^{9-p}q}{(2\pi)^{9-p}}
\,e^{i \vec q \cdot\vec{x}}\,|B_v \rangle \otimes |q_B\rangle
\ee
with $q_B^i = (q^8\,\sin \pi \epsilon, \vec 0, q^8\,\cos \pi \epsilon)$.
The boosted boundary state (in momentum space) $|B_v \rangle$ is obtained by
applying $J^{18}$ to eqs.(\ref{mom})-(\ref{zm}). 
The result corresponds to the replacement \cite{noi}
\bea
M^{ij} &\rightarrow& M^{ij}(v) = (\Sigma(v) M \Sigma^T(v))^{ij} \\
M_{ab} &\rightarrow& M_{ab}(v) = (\Sigma(v) M \Sigma^T(v))_{ab}
\eea
where $\Sigma(v)$ is the appropriate representation of the $SO(8)$ 
rotation:
\beqa
\Sigma_v(v)&=&\pmatrix {\cos \pi \epsilon & 0 &-\sin \pi \epsilon \cr
0 &I_{6}&0\cr
\sin \pi \epsilon & 0 &\cos \pi \epsilon \cr} \\
\Sigma_s(v)&=&\cos (\frac {\pi \epsilon}2)\,\delta_{a b}
-\sin (\frac {\pi \epsilon}2)\,\gamma^{18}_{a b}
\eeqa

After diagonalizing $\gamma^{18}$ with a suitable global $SO(8)$ rotation 
of $S^a_n$ and $\tilde S^a_n$, the cylinder amplitude between two Dp-branes
moving with relative velocity $v$ is found to be 
\bea 
{\cal A}=&&\frac{1}{16}V_p\,(2\pi^2)^{4-p}\int_0^\infty \!dt 
\int\frac{d^{8-p}q}{(2\pi)^{8-p}}\,e^{i \vec q \cdot (\vec{x}-\vec{y})}\,
e^{- \frac {\pi}{2} t q^2} Z_0^F  \nn \\ && \qquad \qquad \qquad \quad \;
\times \frac 1{\sin \pi \epsilon} 
\prod_{n=1}^{\infty}\frac{|1-e^{i \pi \epsilon /2} e^{-2\pi tn}|^8}
{|1-e^{i \pi \epsilon} e^{-2\pi tn}|^2(1-e^{-2\pi tn})^6} 
\label{dynamic}
\eea
The zero mode part $Z_0^F$ no longer vanishes
$$
Z_0^F = \langle B_0,v|B_0\rangle = {\rm Tr}_V {M^T(v)M(0)} - {\rm Tr}_S {M^T(v)M(0)} = 
16 \sin^4 \frac {\pi\epsilon}2
$$
and after the analytic continuation $\epsilon \rightarrow i \epsilon$, the 
final result is
\be
{\cal A} = \frac {V_p}{16 \pi i}\, (2\pi^2)^{4-p} 
\int_{0}^{\infty} \frac {dt}{(2\pi t)^{\frac {8-p}2}} 
e^{- \frac {b^2}{2 \pi t}} 
\frac {\vartheta_1^4(i \frac \epsilon 2|2it)}{\vartheta_1(i \epsilon|2it)}
\frac {\vartheta_1^\prime(0|2it)}{\eta^{12}(2it)}
\ee
which coincides with the well known result of ref.\cite{bach} after using 
the Riemann identity and the modular transformation $t \rightarrow 1/t$.

\subsection{Open string channel}

We compute now the phase-shift from the standard \cite{bach} open channel point
of view in the light-cone gauge. 
The $X^+$ and $X^-$ coordinates are here Neumann, as usual 
(and not Dirichlet as before), and we 
could in principle consider only p-branes with $p \geq 2$, although the final
result is clearly extendable to all branes. 

Similarly to ref.\cite{bach}, the action for two moving branes in the frame 
where one of them is at rest, is given by:
$$
S =-\frac{1}{2\pi}\int \!d^2\sigma\partial_{\alpha}X^i\partial^{\alpha}X^i
+\frac{i}{\pi}\int d^2\sigma\bar{S}\rho^a\partial_a S +
\frac{v}{\pi}\oint_{\sigma=\pi}\!\! d\tau
(X^1\partial_{\sigma}X^8-\frac{i}{4}\bar{S}\rho^1\gamma^{18}S) 
$$
Varying this action, we require the boundary term to vanish and solve for the
constraint; for the bosonic coordinates the result is identical to that of 
ref.\cite{bach} (with $\epsilon\rightarrow i\epsilon$), while the fermionic 
boundary conditions are found imposing 
$\delta S^a=-iM_{ab}\delta\tilde{S}^{b}$
The result is:
\bea 
S^a(\tau,\sigma)&=&P_+^{ab}\sum_{n=-\infty}^{+\infty}S_{-n}^b
e^{-i(n+\epsilon/2)(\tau+\sigma)}+P_-^{ab}\sum_{n=-\infty}^{+\infty}S_{-n}^b
e^{-i(n-\epsilon/2)(\tau+\sigma)} \nonumber \\
\tilde{S}^{a}(\tau,\sigma)&=&-iM_{ab}\left(
P_+^{bc}\sum_{n=-\infty}^{+\infty}S_{-n}^c
e^{-i(n+\epsilon/2)(\tau-\sigma)}+P_-^{bc}\sum_{n=-\infty}^{+\infty}S_{-n}^c
e^{-i(n-\epsilon/2)(\tau-\sigma)}\right) 
\eea
where $P_{\pm}=1/2 (1 \pm i\gamma^{18})$ and $\tan\pi\epsilon=v$.

The fermionic part of the normal ordered light-cone Hamiltonian $P^-$ is then
\bea 
P^-_F&=&\frac{1}{2}\sum_{n=-\infty}^{+\infty}
\left(S_{-n}^a(1+i\gamma^{18})_{ab}
S_n^b(n+\frac{\epsilon}{2})+S_{-n}^a(1- i\gamma^{18})_{ab}
S_n^b(n-\frac{\epsilon}{2})\right)\nonumber \\
&=& \sum_{n=1}^{\infty}(2n\,S_{-n}^a\,S_n^a+\epsilon\, 
S_{-n}^a i\gamma^{18}_{ab} S_n^b)+\frac{\epsilon}{2}\,
S_0^a i\gamma^{18}_{ab}S_0^b 
\eea
Note that since the $S_n$ modes are space-time fermions, they are twisted
by $\epsilon/2$, whereas the bosonic coordinate present a twist of $\epsilon$.
This implies that the eight $\epsilon/2$-twisted fermions and the two 
$\epsilon$-twisted bosons give a total contribution to the Hamiltonian equal to
\be 
8\cdot\frac{1}{4}\frac{\epsilon}{2}(1-\frac{\epsilon}{2})
-2\cdot\frac{1}{4}\epsilon(1-\epsilon)=\frac{\epsilon}{2} 
\ee
As before, we can perform an $SO(8)$ rotation to the $S_{n}$ oscillator modes 
to put the $S_{-n} i\gamma^{18}S_n$ term in a diagonal form; the total Hamiltonian 
$P^-$ takes then the following form:
\bea
P^-=\frac{1}{2p^+}&&\left[(p^i)^2+\,\frac{b^2}{\pi^2}+2\,\sum_{a=1}^{4}
(n+\frac{\epsilon}{2})S^a_{-n}S^a_n+2\,\sum_{a=5}^{8}(n-\frac{\epsilon}{2})
S^a_{-n}S^a_n \right.\nonumber \\ 
&& + \left.\frac{\epsilon}{2}\,S_0^a i\gamma^{18}_{ab}S_0^b +\,\epsilon\,+\,
(\mbox{bos. osc.})\right] 
\eea
We can now compute the partition function
\bea 
{\cal A} &=&\int_0^{\infty}\frac{dt}{t}\,{\rm Tr}\,
e^{-\pi tp^+(P^--i\partial_+)} \nn \\
&=& V_p \int_0^{\infty}\frac{dt}{t}\int\frac{d^pp}{(2\pi)^p}
e^{-\frac{\pi t}{2}(p^2+b^2/\pi^2)}\frac{e^{-\pi t\epsilon/2}}
{1-e^{-\pi t\epsilon}}{\rm Tr}_{S_0}\,e^{-i\pi t\epsilon R_0^{18}} \times
\\ &\;& \qquad \times \prod_{n=1}^{\infty}\frac{(1-e^{-\pi t(n+\epsilon/2)})^4
(1-e^{-\pi t(n-\epsilon/2)})^4}{(1-e^{-\pi tn})^6(1-e^{-\pi(n+\epsilon ) t})
(1-e^{-\pi(n-\epsilon )t})} \nn 
\eea
The trace over the zero modes $S_0$ yields
\be 
{\rm Tr}_{S_0}\,e^{-i\pi t\epsilon R_0^{18}} = 
16 \sinh^4 \frac{\pi t\epsilon}{4} 
\ee
Performing finally the integral over the momentum and the analytic continuation
$\epsilon\rightarrow i\epsilon$, the result can be written in 
terms of $\vartheta$-functions as:
\be 
\label{Aop}
{\cal A}=\frac{V_p}{2\pi i}\int_0^{\infty}\frac{dt}{t}(2\pi^2 t)^{-\frac{p}{2}}
\,e^{-\frac{b^2t}{2\pi}}\,
\frac{\vartheta_1^4(\frac{\epsilon t}{4}|\frac{it}{2})} 
{\vartheta_1(\frac{\epsilon t}{2}|\frac{it}{2})}
\frac {\vartheta_1^{\prime}(0|\frac{it}{2})}{\eta^{12}(\frac{it}{2})} 
\ee
Again, eq.(\ref{Aop}) reproduces the usual result after having
performed the spin-structure sum.


\begin{thebibliography}{99}

\bibitem{pol} J. Polchinski, {\it Dirichlet-Branes and Ramond-Ramond charges},
Phys. Rev. Lett. {\bf 75} (1995) 4724, hep-th/9510017; 
{\it TASI lectures on D-branes}, hep-th/9611050.

\bibitem{bach} C. Bachas, {\it D-brane dynamics}, Phys. Lett. {\bf B374} 
(1996) 37, hep-th/9511043; {\it (Half) a lecture on D-branes}, hep-th/9701019.

\bibitem{dan} U. H. Danielsson, G. Ferretti and B. Sundborg, 
{\it D-particle dynamics and bound states}, 
Int. J. Mod. Phys. {\bf A11} (1996) 5463, hep-th/9603081.

\bibitem{kp} D. Kabat and P. Pouliot, {\it A comment on zero-brane quantum
mechanics}, Phys. Rev. Lett. {\bf 77} (1996) 1004, hep-th/9603127.

\bibitem{lif} G. Lifschytz, {\it Comparing D-branes to black-branes},
Phys. Lett. {\bf B388} (1996) 720, hep-th/9604156.

\bibitem{dkps} M.R. Douglas, D. Kabat, P. Pouliot and S.H. Sphenker, 
{\it D-branes and short distances in string theory}, Nucl. Phys. {\bf B485} 
(1997) 85, hep-th/9608024.

\bibitem{pocai} J. Polchinski and Y. Cai, {\it Consistency of open superstring
amplitudes}, Nucl. Phys. {\bf B296} (1988) 91.

\bibitem{clny} C.G. Callan, C. Lovelace, C.R. Nappi and S.A. Yost, 
{\it Adding holes and crosscaps to the superstring}, 
Nucl. Phys. {\bf B293} (1987) 83; 
{\it Loop corrections to superstring equations of motion}, 
Nucl. Phys. {\bf B308} (1988)  221.

\bibitem{li} M. Li, {\it Boundary states of D-branes and Dy-strings}, 
Nucl. Phys. {\bf B460} (1996) 351, hep-th/9510161.

\bibitem{calkl} C.G. Callan and I.R. Klebanov, 
{\it D-brane boundary state dynamics}, 
Nucl. Phys. {\bf B465} (1996) 473, hep-th/9511173.

\bibitem{billo} M. Bill\`o,  P. Di Vecchia and D. Cangemi, 
{\it Boundary states for  moving D-branes}, Phys. Lett. {\bf B400} (1997) 63,
hep-th/9701190.

\bibitem{lerda1} M. Frau, I. Pesando, S. Sciuto, A. Lerda and R. Russo, 
{\it Scattering of closed strings from many D-branes}, 
Phys. Lett. {\bf B400} (1997) 52, hep-th/9702037.

\bibitem{lerda2} P. Di Vecchia, M. Frau, I. Pesando, S. Sciuto, A. Lerda and 
R. Russo,  {\it Classical p-branes from boundary state}, 
Nucl. Phys. {\bf B507} (1997) 259, hep-th/9707068.

\bibitem{hin} F. Hussain, R. Iengo and C. N\'u\~nez, 
{\it Axion production from gravitons off interacting 0-branes}, 
Nucl. Phys. {\bf B497} (1997) 205, hep-th/9701143.

\bibitem{hins} F. Hussain, R. Iengo, C. N\'u\~nez and C.A. Scrucca, 
{\it Interaction of moving D-branes on orbifolds}, 
Phys. Lett. {\bf B409} (1997) 101, hep-th/9706186; 
{\it Closed string radiation from moving D-branes}, Nucl. Phys. {\bf B517} 
(1998) 92, hep-th/9710049.

\bibitem{bis} M. Bertolini, R. Iengo and C.A. Scrucca,
{\it Electric and magnetic interaction of dyonic D-branes and odd spin 
structure}, Nucl. Phys. {\bf B522} (1998) 193, hep-th/9801110.

\bibitem{gre} M.B. Green, {\it Point-like states for type IIB superstrings}, 
Phys. Lett. {\bf B329} (1994) 435, hep-th/9403040.

\bibitem{gregut1} M.B. Green and M. Gutperle, {\it Light-cone supersymmetry
and D-branes}, Nucl. Phys. {\bf B476} (1996) 484, hep-th/9604091.

\bibitem{gregut2} M.B. Green and M. Gutperle, {\it Effects of D-instantons}, 
Nucl. Phys. {\bf B498} (1997) 195, hep-th/9701093.

\bibitem{barb} J.L.F. Barb\'on, {\it Fermion exchange between D-instantons}, 
Phys. Lett. {\bf B404} (1997) 33, hep-th/9701075.

\bibitem{bfss} T. Banks, W. Fischler, S. Shenker and L. Susskind, 
{\it M-theory as a matrix model: a conjecture}, Phys. Rev. {\bf D55} (1997) 112,
hep-th/9610043.

\bibitem{harv} J.A. Harvey, {\it Spin dependence of D0-brane interactions},
Nucl. Phys. Proc. Suppl. {\bf B68} (1998) 113, hep-th/9706039.

\bibitem{noi} J.F. Morales, C.A. Scrucca and M. Serone, 
{\it A note on supersymmetric D-brane dynamics}, Phys. Lett. {\bf B417}
(1998) 233, hep-th/9709063.

\bibitem{kra} P. Kraus, {\it Spin-orbit interaction from matrix theory}, 
Phys. Lett. {\bf B419} (1998) 73, hep-th/9709199.

\bibitem{gsw} M.B. Green, J.H. Schwarz and E. Witten, 
{\it Superstring theory: volume 2}, Cambridge University Press 1987.

\bibitem{witten} E. Witten, {\it String theory dynamics in various dimensions},
Nucl. Phys. {\bf B443} (1995) 85, hep-th/9503124.

\bibitem{douglas} M. Berkooz and M.R. Douglas, {\it Five-branes in M(atrix)
theory}, Phys. Lett. {\bf B395} (1997) 196, hep-th/9610236.

\bibitem{aich} P.C. Aichelburg and F. Embacher, {\it Exact superpartners of N=2
supergravity solitons}, Phys. Rev. {\bf D34} (1986) 3006.

\bibitem{dlrbig} M.J. Duff, J.T. Liu and J. Rahmfeld, {\it Dipole moments of black
holes and string states}, Nucl. Phys. {\bf B494} (1997) 161, hep-th/9612015.

\bibitem{dlr} M.J. Duff, J.T. Liu and J. Rahmfeld, {\it g=1 for Dirichlet 0-branes}, Nucl. Phys. {\bf B524} (1998) 129, hep-th/9801072.

\bibitem{hs} G.T. Horowitz and A. Strominger, {\it Black strings and p-branes},
Nucl. Phys. {\bf B360} (1991) 197.

\bibitem{mp} R.C. Myers and M.J. Perry, {\it Black holes in higher dimensional
space-times}, Ann. Phys. {\bf 172} (1986) 304.

\bibitem{hioy} A. Hosoya, K. Ishikawa, Y. Ohkuwa and K. Yamagishi, {\it Gyromagnetic ratio
of heavy particles in the Kaluza-Klein theory}, Phys. Lett. {\bf B134} (1984) 44.

\bibitem{wei} S. Weinberg, {\it Dynamic and algebraic symmetries}, in {\it Lectures on
Elementary Particles and Quantum Field Theory}, (MIT Press, Cambridge, 1970).

\bibitem{fpt} S. Ferrara, M. Porrati and V.L. Telegdi, {\it g=2 as the natural value of the
tree-level gyromagnetic ratio of elementary particles}, Phys. Rev. {\bf D46} (1992) 3529.

\bibitem{jack} R. Jackiw, {\it g=2 as a gauge condition}, Phys. Rev. {\bf D57} (1998) 2635,
hep-th/9708097.

\bibitem{AB} O. Aharony and M. Berkooz, {\it Membrane dynamics in M(atrix) theory},
Nucl. Phys. {\bf B491} (1997) 184, hep-th/9611215.

\bibitem{lm} G. Lifschytz and S. D. Mathur, {\it Supersymmetry 
and Membrane Interactions in M(atrix) theory}, Nucl. Phys. {\bf B507} 
(1997) 621, hep-th/9612087.

\bibitem{lifs} G. Lifschytz, {\it Four-brane and six-brane interactions in M(atrix) theory},
hep-th/9612223; {\it A note on the transverse five-brane in M(atrix) theory},
Phys. Lett. {\bf B409} (1997) 124, hep-th/9703201.

\bibitem{tse} I. Chepelev and A. A. Tseytlin,
{\it  Long-distance interactions of D-brane bound states and longitudinal 5-brane in 
M(atrix) theory}, Phys. Rev. {\bf D56} (1997) 3672, hep-th 9704127; 
{\it Interactions of type IIB D-branes from D-instanton matrix model },
hep-th/9705120, Nucl. Phys. {\bf B511} (1998) 629.

\bibitem{pierre} J.M. Pierre, {\it Interactions of eight-branes in string theory and 
M(atrix) theory}, Phys. Rev. {\bf D57} (1998) 1250, hep-th/9705110.

\bibitem{bb} K. Becker and M. Becker, {\it A two-loop test of M(atrix) theory},
Nucl. Phys. {\bf B506} (1997) 48, hep-th/9705091.

\bibitem{bbpt} K. Becker, M. Becker, J. Polchinski and A. Tseytlin, 
{\it Higher order graviton scattering in M(atrix) theory}, Phys. Rev. {\bf D56}
(1997) 3174, hep-th/9706072.

\end{thebibliography}
\end{document}